\documentclass[aps,preprint,pra,superscriptaddress,longbibliography,nofootinbib,notitlepage]{revtex4-1}

\usepackage[letterpaper, margin=1in]{geometry}
\usepackage{amsmath,amsfonts,amssymb}
\usepackage{braket}
\usepackage{graphicx}
\usepackage{tikz}
\usetikzlibrary{shapes,arrows,positioning,calc}
\usepackage[hidelinks]{hyperref}
\usepackage{slashed}
\usepackage{setspace}

\newcommand{\mb}[1]{{\mathbf #1}}

\DeclareMathOperator{\tr}{tr}

\newcommand{\be}{\begin{equation}}
\newcommand{\ee}{\end{equation}}

\newcommand{\Oi}{\mathcal{O}}

\newcommand{\ul}{\underline}

\AtBeginDocument{%
    \newwrite\bibnotes
    \def\bibnotesext{Notes.bib}
    \immediate\openout\bibnotes=\jobname\bibnotesext
    \immediate\write\bibnotes{@CONTROL{REVTEX41Control}}
    \immediate\write\bibnotes{@CONTROL{%
    apsrev41Control,author="08",editor="1",pages="1",title="0",year="1"}}
     \if@filesw
     \immediate\write\@auxout{\string\citation{apsrev41Control}}%
    \fi
}%

\begin{document}

\title{Classical-quantum scattering}

\author{Daniel Carney}
\email{carney@lbl.gov}
\affiliation{Physics Division, Lawrence Berkeley National Laboratory, Berkeley, CA 94720, USA}
\author{Akira Matsumura}
\email{matsumura.akira@phys.kyushu-u.ac.jp}
\affiliation{Department of Physics, Kyushu University, Fukuoka, 819-0395, Japan}

\date{\today}

\begin{abstract}
We analyze the framework recently proposed by Oppenheim \emph{et al.} to model relativistic quantum fields coupled to relativistic, classical, stochastic fields (in particular, as a model of quantum matter coupled to ``classical gravity''). Perhaps surprisingly, we find that we can define and calculate scattering probabilities which are Lorentz-covariant and conserve total probability, at least at tree level. As a concrete example, we analyze $2 \to 2$ scattering of quantum matter mediated by a classical Yukawa field. Mapping this to a gravitational coupling in the non-relativistic limit, and assuming that we can treat large objects as point masses, we find that the simplest possible ``classical-quantum'' gravity theory constructed this way gives predictions for $2 \to 2$ gravitational scattering which are inconsistent with simple observations of, e.g., spacecraft undergoing slingshot maneuvers. We comment on lessons learned for attempts to couple quantum matter to ``non-quantum'' gravity, or more generally, for attempts to couple relativistic quantum and classical systems.

\end{abstract}

\maketitle

\newpage

\renewcommand{\baselinestretch}{1.0}\normalsize
\tableofcontents
\renewcommand{\baselinestretch}{1.25}\normalsize

\section{Introduction and motivation}

When the curvature of spacetime is small compared to the Planck scale $R/M_{\rm pl}^2 \ll 1$, gravity can be quantized perturbatively as an effective quantum field theory, with action
\be
\label{eq:action}
S = M_{\rm pl}^2 \int d^4x \sqrt{-g} \left( R + \mathcal{L}_{\rm matter} + \frac{c_1}{M_{\rm pl}^2} R^2 + \frac{c_2}{M_{\rm pl}^2} R_{\mu\nu}R^{\mu\nu} + \cdots \right)
\ee
where the dots represent an infinite series in powers of curvature in Planck units~\cite{Donoghue:1995cz,burgess2004quantum,Donoghue:2022eay}. The theory is non-renormalizable: when $R/M_{\rm pl}^2 \gtrsim 1$, this infinite series of terms becomes dominant over the Einstein-Hilbert term $R$. Because the coefficients $c_i$ are unknown, the theory is non-predictive in this regime. However, in a laboratory experiment involving solid objects, $R/M_{\rm pl}^2 \approx 10^{-91}$, so only the Einstein-Hilbert term contributes meaningfully. Thus, this effective quantum field theory treatment of the weak gravitational field is extremely well-defined in settings other than near a singularity, such as the deep interior of a black hole or the very early universe.

While quantizing gravity this way is theoretically self-consistent at ordinary scales, it is interesting to consider the possibility that gravity is not actually quantized this way in nature. Investigating this notion is particularly motivated by the possibility of experimental tests that could distinguish standard quantized gravity from alternative models~\cite{Carney:2018ofe,Lindner:2004bw,Kafri:2013wxa,bose2017spin,marletto2017gravitationally,Matsumura:2020law,Carney:2021yfw,Datta:2021ywm,Oppenheim:2022xjr,Lami:2023gmz,Carney:2023nzz,Carney:2024dsj}. Recently, a number of models in which the gravitational field is ``classical'' have been constructed to make this suggestion more concrete~\cite{Kafri:2014zsa,Penrose:1996,tilloy2016sourcing,Grossardt:2022zsi,Carney:2023aab}. Generally, these models have been in the non-relativistic regime, and ``classical'' means that the Newtonian interaction cannot generate entanglement between two massive systems. This would be in stark contrast to standard quantized gravity as in Eq. \eqref{eq:action}, which definitely predicts entanglement generation through virtual graviton exchange, or equivalently, through the Newtonian two-body operator obtained when integrating out the gravitons in the low-velocity limit~\cite{Carney:2021vvt}.

Very recently, Oppenheim and collaborators have suggested an interesting relativistic generalization of these ``classical'' gravity models~\cite{Oppenheim:2018igd,Oppenheim:2023izn,Layton:2023oud,Weller-Davies:2024mea,Grudka:2024llq}. They refer to their general setting as ``classical-quantum'' (CQ) gravity and we will follow this terminology. In their CQ model, the gravitational field is a bone fide classical field. It undergoes stochastic time evolution and is coupled to quantum matter in a way made precise below. It is very interesting to determine if this kind of CQ framework is mathematically self-consistent, and if it can be used to formulate a physically viable model of gravity. Our purpose in this paper is to make some steps towards answering these questions.

The general CQ framework is independent of gravity and we choose to specialize to the simpler case of a Yukawa model, where a classical Yukawa field $\phi$ couples to some quantized matter, which avoids any complications coming from gauge symmetry. On the other hand, for non-relativistic questions, a Yukawa interaction in the massless limit generates a $1/r$ potential, so it can be used to analyze the Newtonian limit of gravity by identifying the coupling constants appropriately.

As a straightforward observable, we attempt to formulate scattering theory for quantum matter coupled through this CQ Yukawa interaction. This proves to be instructive in many ways. For one thing, it raises some basic concerns about the purely diffusive evolution law assumed for the classical field~\cite{Oppenheim:2023izn,Weller-Davies:2024mea}, which makes it hard to even formulate the scattering problem because the theory does not have a stable ground state.\footnote{Penington~\cite{geoff} has estimated that, in the context of gravity, the stochastic generation of gravitational waves from this diffusive evolution is quantitatively inconsistent with non-observation of a stochastic gravitational wave background in LIGO~\cite{LIGOScientific:2016jlg}.} Nevertheless, allowing for some choices and approximations which we make clear, we formulate the $2 \to 2$ scattering problem for two massive bodies. The CQ framework has two free functions, called $D_0$ and $D_2$, and we analyze only the simplest case where these are assumed to be constants. We find that noise effects in the resulting CQ model produce large deviations in the scattering distributions of even non-relativistic particles. Once the gravitational parameters are inserted, the CQ model predicts effects which are in direct conflict with observations. We discuss the generality of this conclusion in Sec. \ref{sec-conclusions}.

Beyond the specifics of these gravity models, it is of substantial interest to understand if classical systems can be coupled to quantum ones in some way compatible with special relativity, let alone general relativity~\cite{banks1984difficulties,unruh1995evolution,diosi2022there,matsumura2023reduced}. One might hope that the quantitative difficulties found here come more from specific choices in the construction rather than the overall idea of a relativistic framework for coupling classical systems to quantum ones. Conversely, this construction appears perhaps more robust than one might have anticipated. For example, it appears capable of producing Lorentz-covariant scattering probabilities, at least at leading order in a perturbative expansion. We end with a few comments on this, with the goal of deriving some general lessons for possible relativistic classical-quantum evolution laws.

\section{Classical-quantum framework}
\label{sec:cq}

To motivate the general idea of coupling quantum systems to classical ones, and to highlight some of the features that arise in the model of Oppenheim \emph{et al.}, consider a quantum computer operating under quantum error correction. This fundamentally consists of quantum systems (qubits) which are periodically being measured (to determine the error syndromes). These measurements are done by a classical apparatus: some readout system which ultimately takes qubit states and records measurement outcomes as classical data. This classical data is then used to determine the appropriate feedback unitary that needs to be applied to the qubits. This feedback unitary is also implemented by an effectively classical system, like a control laser or microwave drive. Thus, the qubits and classical control system together constitute a dynamically coupled quantum-classical system.

\begin{figure}[t]

\begin{tikzpicture}[scale=0.5]

\draw (2,3) -- (4,3);
\node at (1,3) {$\ket{\psi_1}$};

\draw (2,1) -- (4,1);
\node at (1,1) {$\ket{0}$};

\draw (2,-1) -- (4,-1);
\node at (1,-1) {$\ket{0}$};

\draw (2,-3) -- (4,-3);
\node at (1,-3) {$\ket{\psi_2}$};

\draw [black] (4,0.5) rectangle (6,3.5);
\node at (5,2) {$U_{\rm ent}$};

\draw [black] (4,-0.5) rectangle (6,-3.5);
\node at (5,-2) {$U_{\rm ent}$};

\draw (6,1) -- (8,1);
\draw [black] (8,1.5) -- (8,.5) arc(-90:90:.5) -- cycle;

\draw (6,-1) -- (8,-1);
\draw [black] (8,-1.5) -- (8,-.5) arc(90:-90:.5) -- cycle;

\draw (6,3) -- (11,3);
\draw [black] (11,2.5) rectangle (13,3.5);
\node at (12,3) {$U_{\rm FB}$};

\draw (6,-3) -- (11,-3);
\draw [black] (11,-2.5) rectangle (13,-3.5);
\node at (12,-3) {$U_{\rm FB}$};

\draw [double] (8.5,1) -- (11,-2.5);
\draw [double] (8.5,-1) -- (11,2.5);

\draw (13,3) -- (14.5,3);
\draw [dashed] (8.5,1) -- (14.5,1);

\draw (13,-3) -- (14.5,-3);
\draw [dashed] (8.5,-1) -- (14.5,-1);

\end{tikzpicture}

\caption{Measurement-and-feedback gravity~\cite{Kafri:2014zsa,Carney:2023aab}. Shown here is one timestep in a Markovian process describing the gravitational interaction of two massive systems $i=1,2$. Each system is coupled to an ancilla, which performs a position measurement to obtain an estimate of the two positions $\braket{\mb{x}_i}$. This information is then fed back through a one-body operator $U_{\rm FB} = e^{i V_{\rm sc} \Delta t}$, with $V_{\rm sc} = G_N m_1 m_2/\left|\mb{x}_i - \braket{\mb{x}_j} \right|$. This generates a noisy open channel on the two masses which reproduces semiclassical Newtonian gravity. Figure reproduced from~\cite{Carney:2018ofe}.}
\label{fig:circuit}
\end{figure}
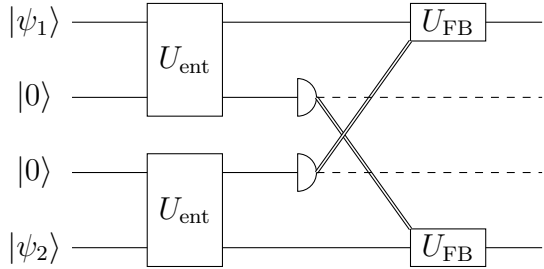

This picture has already been used to construct models of non-relativistic matter coupled through ``semi-classical'' gravitational interactions \cite{Kafri:2014zsa}. See. Fig. \ref{fig:circuit}. These models have the property that they induce noise on the quantum system, coming from a combination of the measurements being done (which have random outcomes as usual in quantum mechanics) and the feedback (which has noise due to the random nature of the control data). In these models, however, the classical system is implicit; it is taken for granted that something is doing the classical feedback control, but this is not invoked as an explicit degree of freedom.

Oppenheim \emph{et al}.'s proposal is essentially to extend this paradigm by introducing a classical field (e.g., the gravitational field) as an explicit dynamical system which performs the readout and control. This field has noise for the same reasons just discussed. In the next subsections, we briefly review this construction.

\subsection{Classical-quantum states}

A quantum system is generally described by a density matrix, while a classical stochastic system is described by a classical probability distribution. Let $q$ denote the canonical position variables of a quantum system, and $Q, P$ some complete set of canonical classical variables on a classical system. In practice these will be position and momentum type variables as indicated by the $q,Q,P$ notation. We are using one symbol for each but the generalization to many degrees of freedom (e.g., $q \to q_1, q_2, \ldots$) is completely straightforward. The basic state variable in CQ theory is an object encoding the state of both of these:
\be
\varrho(Q,P) := \int dq d\ul{q} \, \varrho(q, \ul{q}; Q,P) \ket{q} \bra{\ul{q}}.
\ee
We are writing this as an density matrix-valued function of the classical variables. 

The way to interpret this is as follows. For a fixed classical configuration $(Q, P)$, this object is the density matrix of the quantum system, conditioned on that classical configuration. If we are not measuring the classical system at all, we can also find the reduced state after averaging over the classical system,
\be
\label{eq:rho-qu}
\rho := \int dQ dP \, \varrho(Q,P).
\ee
Conversely, the state (probability distribution) of the classical variables can similarly be considered as either conditioned on or averaged over the quantum system. The probability distribution for the classical variables, conditioned on the quantum system being in state $\ket{q}$, is
\be
\mathcal{P}(Q,P | q_{i}) = \varrho(q, q; Q,P).
\ee
If we are not measuring the quantum system, the reduced state of the classical variable is just this with an average (trace) over the quantum state:
\be
\mathcal{P}(Q,P) = \int dq \, \varrho(q, q; Q,P).
\ee
In what follows we will be particularly interested in the reduced state of just the quantum matter, with the classical degrees of freedom traced out, as in Eq. \eqref{eq:rho-qu}.

\subsection{Time evolution of the full CQ state}

The CQ evolution law is defined by a path integral of the following form:
\be
\varrho(q_f,\ul{q}_f,Q_f,P_f,t_f) = \int Dq D \ul{q} DQ DP e^{I_{CQ}[q,\ul{q},Q,P]} \varrho(q_0,\ul{q}_0,Q_0,P_0,t_0),
\ee
with the boundary conditions set as usual by the arguments on the left-hand side. We will give the form of $I_{CQ}$ shortly. First let us motivate it by considering, separately, the path integral for a quantum system and that for a stochastic classical system.

First consider a quantum system. The density matrix obeys
\be
\dot{\rho} = -i[H,\rho].
\ee
Assuming $H$ is at most quadratic in momenta, this equation famously admits a path integral solution (a particular ``Schwinger-Keldysh'' path integral)
\be
\rho(q_f,\ul{q}_f,t_f) = \int Dq D\ul{q} e^{i (S[q] - S[\ul{q}])} \rho(q_0,\ul{q}_0,t_0).
\ee
Here, the first exponential evolves the ket, while the second evolves the bra, in the sense of $\rho \to U \rho U^\dag$. The action $S$ is defined as usual. The generalization beyond theories quadratic in momenta is standard, but we will not need it in what follows.

Now consider a classical system undergoing stochastic evolution. As a concrete example, consider a particle satisfying the simple Fokker-Planck equation
\be
\dot{\mathcal{P}} = -\frac{P}{m} \frac{\partial}{\partial Q} \mathcal{P} + \frac{D_2}{2} \frac{\partial^2}{\partial P^2} \mathcal{P},
\ee
where $\mathcal{P} = \mathcal{P}(Q,P)$ is the probability distribution on $(Q,P)$. This equation describes pure diffusion in the position variable $Q$, where $D_2$ is a coefficient with units of $M^3$ that sets the rate of diffusion. This equation similarly admits a path integral solution~\cite{risken1996fokker},
\be
\label{eq:FP-example}
\mathcal{P}(Q_f,P_f,t_f) = N \int DQ DP \, \delta\left[\dot{Q}-\frac{P}{m}\right] e^{-\frac{1}{D_2} \int dt \, [m\ddot{Q}]^2} \mathcal{P}(Q_0,P_0,t_0),
\ee
where $N$ is a normalization factor. A good way to understand what this path integral does is as follows.\footnote{We thank Shivaji Sondhi for explaining this to us.} The Fokker-Planck equation can equivalently be described by the Langevin equations of motion for a particle being stochastically kicked,
\be
\label{eq:langevin-example}
\dot{Q} = \frac{P}{m}, \ \ \dot{P} = \sqrt{D_2} \xi \implies \ddot{Q} = \frac{\sqrt{D_2}}{m} \xi,
\ee
where $\xi = \xi(t)$ is a Gaussian-distributed random white noise variable
\be
\mathcal{P}[\xi(t)] = N e^{- \int dt \, \xi^2(t)}.
\ee
Thus, the probability distribution of a given particle configuration $Q_f$ at $t_f$ can be obtained by just solving $\xi = m \ddot{Q}/\sqrt{D_2}$ and inserting it into this distribution, yielding
\be
P(Q_f,t_f ) = N \int DQ e^{- \frac{1}{D_2} \int dt \ \left[ m \ddot{Q} \right]^2} P(Q_0,t_0).
\ee
Notice that this path integral is in real (Lorentzian) time. It is very different from the usual quantum form: the equations of motion $m \ddot{Q} = 0$ (averaged over the Gaussian noise $\braket{\xi} = 0$) appear explicitly in the integral weight. In this sense the path integral represents time evolution on the space of configurations, with the dominant contribution from the $\xi = 0$ solution, and contributions away from this average solution distributed according to the noise $\xi \neq 0$. Note that this path integral is quartic in derivatives, but unlike a quantum-mechanical path integral this does not imply anything about acausal behavior.

Putting these ideas together, Oppenheim \emph{et al}. define a total CQ path integral by evolving the quantum parts with a Schwinger-Keldysh path integral and the classical part with a Fokker-Planck path integral. They then make an ansatz for coupling terms. We refer the reader to~\cite{Oppenheim:2018igd,Oppenheim:2023izn,Layton:2023oud,Weller-Davies:2024mea,Grudka:2024llq} for the general construction. Here we will consider quantum matter, which we model as a pair of distinguishable scalar fields $\chi_{1,2}$, with Yukawa couplings to a classical scalar field $\phi$. The CQ action for this model is
\begin{align}
\begin{split}
 I_{CQ}[\boldsymbol{\chi}, \underline{\boldsymbol{\chi}}, \phi,\pi] & = i(S_0 [\boldsymbol{\chi}]+S_\text{int}[\boldsymbol{\chi},\phi]-S_0 [\underline{\boldsymbol{\chi}}]-S_\text{int}[\underline{\boldsymbol{\chi}},\phi]) +\ln(\delta[\dot{\phi}-\pi])\\
& -\frac{D_0}{2} \int^{t_f}_{t_0} d^4x \left[ \frac{\delta S_\text{int}}{\delta \phi}[\boldsymbol{\chi},\phi]- \frac{\delta S_\text{int}}{\delta \phi}[\underline{\boldsymbol{\chi}},\phi]\right]^2 \\
&
-\frac{1}{2D_2} \int^{t_f}_{t_0} d^4x \left(\dot{\pi}-(\nabla^2-m_{\phi}^2) \phi-\frac{1}{2} \frac{\delta S_\text{int}}{\delta \phi}[\boldsymbol{\chi},\phi]-\frac{1}{2} \frac{\delta S_\text{int}}{\delta \phi}[\underline{\boldsymbol{\chi}},\phi]\right)^2.
\label{eq:I}
\end{split}
\end{align}
Here $\boldsymbol{\chi}=(\chi_1, \chi_2)$ is a multiplet of the two scalars, and $\pi$ is the canonical momentum for the classical $\phi$ field. The log term enforces the canonical variable delta function like that appearing in Eq. \eqref{eq:FP-example}.

Let us describe the role of each term in Eq. \eqref{eq:I}. In the first line, $S_0$ is just the usual kinetic term for the quantum scalar fields. The term
\be
S_{\rm int} = \lambda \int d^4x \, \phi \boldsymbol{\chi}^2
\label{eq:Yukawa}
\ee
is an ordinary Yukawa coupling. Since we are using scalar fields, $\lambda$ has dimensions of mass. The second line, which involves
\be
\frac{\delta S_{\rm int}}{\delta \phi} = \lambda \boldsymbol{\chi}^2
\ee
generates decoherence of the quantum matter. Specifically, in the field configuration basis $\ket{\boldsymbol{\chi}}$, this term causes off-diagonal terms in the density matrix (i.e., the elements $\ket{\boldsymbol{\chi}} \bra{\ul{\boldsymbol{\chi}}}$ with $\boldsymbol{\chi} \neq \ul{\boldsymbol{\chi}}$) to decay at a rate proportional to $D_0 \lambda^2$. Here $D_0$ is called the decoherence parameter and in this model has units of $D_0 \sim 1/M^2$. 

Finally, the third line generates the diffusive, stochastic evolution of the classical field $\phi$. The overall coefficient $D_2$, called the diffusion parameter, has units of $D_2 \sim M^2$ in this model. Just like the Brownian particle example above, the diffusion is such that the most likely path satisfies a classical equation of motion without noise, and then there are fluctuations around this. The highly novel part is that the classical $\phi$ equations of motion are sourced by the \emph{quantum} field, through the last terms. Essentially, the classical fields become sourced by expectation values of the quantum fields. In this example, we have $\braket{\dot{\pi}} = [\nabla^2 - m_{\phi}^2] \braket{\phi} + \lambda \braket{\boldsymbol{\chi}^2} + \sqrt{D_2/2} \braket{\xi}$. We explain this in detail in Appendix \ref{app:C}. What this equation shows is that the classical fields are sourced by \emph{two} systems: the quantum field, and some implicit external driving field which exerts a white noise force.

The construction given above is supposed to have three key features~\cite{Oppenheim:2018igd,Oppenheim:2023izn,Layton:2023oud,Weller-Davies:2024mea,Grudka:2024llq}. The obvious one is that it gives a coupling of quantum matter to classical matter. Another is that the time evolution generated by this path integral maps density matrices to density matrices and in particular preserves total probability for the whole system. For this to hold, one requires~\cite{Oppenheim:2018igd,Oppenheim:2023izn,Layton:2023oud,Weller-Davies:2024mea,Grudka:2024llq}.
\be
D_0 D_2 \geq 1,
\ee
which Oppenheim \emph{et al.} refer to as a ``decoherence-diffusion tradeoff''. Finally, this path integral is supposed to be Lorentz covariant, and in the gravitational context, can be made generally covariant. In general, $D_{0,2} = D_{0,2}(q,Q,P)$ are allowed to be functions of the canonical path integral variables. In this paper, we are only analyzing the simplest case where these are taken to be simple constants.

The meaning of Lorentz covariance in this framework deserves some discussion. It is clear that the action here is Lorentz invariant in the simple sense of having no free spacetime indices. What is less clear is how to define Lorentz or general coordinate transformations on the CQ states. Since Oppenheim \emph{et al.} have not given a prescription for this, we will not generally attempt it here. However, in the context of scattering theory in Sec. \ref{sec:scattering}, we will define Lorentz transformation properties of the asymptotic states for the quantum matter using the standard rules of quantum field theory. Interestingly, we will find that the scattering probabilities constructed with these states are Lorentz invariant in the sense of being given by Lorentz-invariant functions. However, there is an important twist: the probabilities violate energy-momentum conservation. These two facts are consistent because this model is fundamentally an open quantum system, and thus there is no Noether's theorem~\cite{matsumura2023reduced,marvian2014extending}. Concretely, energy and momentum can be injected both by the classical field, and by the white noise $\xi$ driving the classical field.

\subsection{Time evolution of the reduced quantum state}
\label{sec:reduced}

The CQ path integral generates time-local dynamics on the joint classical and quantum system. However, if we are only interested in the dynamics of the quantum system, we can obtain a non-local path integral by simply doing the integration over the classical system. The result is the standard Feynman-Vernon path integral, representing an open quantum system~\cite{feynman2000theory,caldeira1983path}. This will be particularly useful for us in order to study scattering of quantum matter, because the path integral constructed this way averages over the final configuration of the classical field, which we will assume is unmeasured in the examples below.

More concretely, assume that the joint CQ state at a time $t_0$ has a product form $\varrho(q_0, \ul{q}_0, Q_0,P_0,t_0)=\rho(q_0,\ul{q}_0,t_0) \mathcal{P}(Q_0,P_0,t_0)$. The time evolution of the quantum system at a later time $t_f$ is given by the Feynman-Vernon path integral 
\be
\rho(q_f,\ul{q}_f,t_f) = \int Dq D \ul{q} \, e^{iS[q]-iS_0[\ul{q}] + iS_{IF}[q, \ul{q}]}\rho(q_0,\ul{q}_0,t_0) 
\ee
with the influence functional $S_\text{IF}[q,\ul{q}]$ defined through the following equation,
\be
e^{iS_0 [q]-iS_0[\ul{q}] + iS_{IF}[q, \ul{q}]}= \int DQ DP \, e^{I_{CQ}[q,\ul{q},Q,P]} \mathcal{P}(Q_0,P_0,t_0).
\ee
Here $S_0 [q]$ is again the free action of quantum system (which in a more general model would include any self-interactions). The influence functional $S_{IF}[q,\ul{q}]$ contains all effects from the classical system coupled with the quantum system of interest. We give explicit computations for $S_{IF}$ in our Yukawa model in Appendix \ref{app:fv}.

\section{Classical-quantum scattering}
\label{sec:scattering}

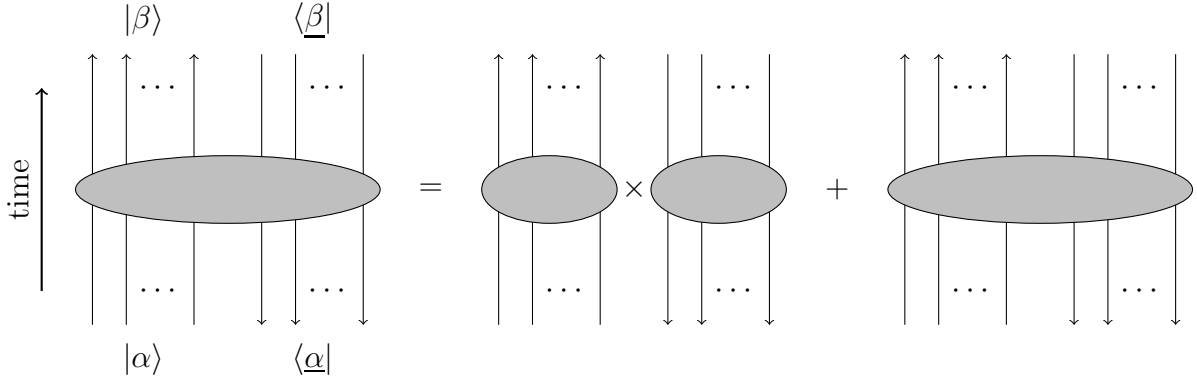
\begin{figure}[t]

\begin{tikzpicture}[scale=.45]


\draw [->,thick] (-5.5,-3) -- (-5.5,3);
\node [rotate=90] at (-6.2,0) {time};

\draw [->] (-4,-4) -- (-4,4);
\draw [->] (-3,-4) -- (-3,4);
\node at (-2,-3) {$\cdots$};
\node at (-2,3) {$\cdots$};
\draw [->] (-1,-4) -- (-1,4);

\node at (-2.5,-5) {$\ket{\alpha}$};
\node at (-2.5,5) {$\ket{\beta}$};

\draw [<-] (1,-4) -- (1,4);
\draw [<-] (2,-4) -- (2,4);
\node at (3,-3) {$\cdots$};
\node at (3,3) {$\cdots$};
\draw [<-] (4,-4) -- (4,4);

\node at (2.5,-5) {$\bra{\ul{\alpha}}$};
\node at (2.5,5) {$\bra{\ul{\beta}}$};

\draw [fill=lightgray] (0,0) ellipse [x radius=4.5, y radius=1];

\node at (6,0) {$=$};


\begin{scope}[shift={(12,0)}]

\draw [->] (-4,-4) -- (-4,4);
\draw [->] (-3,-4) -- (-3,4);
\node at (-2,-3) {$\cdots$};
\node at (-2,3) {$\cdots$};
\draw [->] (-1,-4) -- (-1,4);

\draw [fill=lightgray] (-2.5,0) ellipse [x radius=2, y radius=1];

\node at (0,0) {$\times$};

\draw [<-] (1,-4) -- (1,4);
\draw [<-] (2,-4) -- (2,4);
\node at (3,-3) {$\cdots$};
\node at (3,3) {$\cdots$};
\draw [<-] (4,-4) -- (4,4);

\draw [fill=lightgray] (2.5,0) ellipse [x radius=2, y radius=1];

\node at (6,0) {$+$};

\end{scope}


\begin{scope}[shift={(24,0)}]

\draw [->] (-4,-4) -- (-4,4);
\draw [->] (-3,-4) -- (-3,4);
\node at (-2,-3) {$\cdots$};
\node at (-2,3) {$\cdots$};
\draw [->] (-1,-4) -- (-1,4);

\draw [<-] (1,-4) -- (1,4);
\draw [<-] (2,-4) -- (2,4);
\node at (3,-3) {$\cdots$};
\node at (3,3) {$\cdots$};
\draw [<-] (4,-4) -- (4,4);

\draw [fill=lightgray] (0,0) ellipse [x radius=4.5, y radius=1];

\end{scope}

\end{tikzpicture}

\caption{Schwinger-Keldysh-Feynman-Vernon diagrams for CQ theories. Here we show a general process with many incoming particles in both the ket $\ket{\alpha}$ and bra $\ket{\ul{\alpha}}$ variables. Physical time runs from bottom to top and the arrows represent phase time evolution as either forward or ``backward'' (i.e., like $e^{+i H t}$ acting on the bra). On the right hand side, the first diagram represents time evolution in a closed quantum system: the ket and bra lines talk to themselves but not to each other. The last term is unique to open systems like the CQ model studied here, which allows for correlations to develop between the ket and bra variables.}
\label{fig:scattering}

\end{figure}

Scattering theory in quantum mechanics and quantum field theory is defined in terms of states of well-separated particles freely propagating in from $t \to -\infty$, interacting near $t=0$, and then evolving into (superpositions of) states of well-separated particles propagating freely out to $t \to +\infty$. We will attempt to do the same in our CQ Yukawa model, but this will require a few assumptions and modifications to the usual quantum treatment.

In unitary quantum mechanics, one usually calculates a scattering matrix $S_{\alpha \to \beta} = \braket{\beta | U | \alpha}$, where $U$ generates time evolution from past to future infinity. Here $\alpha = \mb{p}_1,\sigma_1, \mb{p}_2,\sigma_2, \ldots$ are the quantum numbers of the initial state, and $\beta = \mb{p}_1', \sigma_1',\mb{p}_2',\sigma_2', \ldots$ are the quantum numbers of the final state, in the notation of~\cite{weinberg2015lectures,weinberg1995quantum}. The probability of transition $\ket{\alpha} \to \ket{\beta}$ is then given by
\be
P(\alpha \to \beta) \sim |S_{\alpha \to \beta}|^2,
\ee
where the overall proportionality involves correctly normalizing the states to a flux of incoming particles, as we discuss below. 

In CQ theory, there are two twists to this story. One is that the asymptotic states now also need to include the probability distribution of the classical field at past and future infinity. The other is that the model is fundamentally an open system, and thus there is no unitary $S$-matrix. Instead, one has to work directly with probabilities. To circumvent the first issue, here we are going to just assume a certain initial state for the classical field---namely, the distribution with the field completely turned off $\mathcal{P}[\phi,\pi] = \delta[\phi] \delta[\pi]$---and then average over the final $\phi$ states by using a Feynman-Vernon path integral (see Sec. \ref{sec:reduced}). This will be used to compute what we will call a scattering supermatrix element
\be
\label{eq:slashM}
\slashed{M}_{\alpha \ul{\alpha} \to \beta \ul{\beta}} = \braket{\beta | \mathcal{M} \left[ \ket{\alpha} \bra{\ul{\alpha}} \right] | \ul{\beta}},
\ee
where $\mathcal{M}$ is the channel on the quantum system generated by the Feynman-Vernon path integral. From $\slashed{M}$ one obtains the probability for a transition $\ket{\alpha} \to \ket{\beta}$ by
\be
\label{eq:Pgeneral}
P(\alpha \to \beta) \sim \slashed{M}_{\alpha \alpha \to \beta \beta}.
\ee
We will give a much clearer and more explicit formula below in the examples. The proportionality factor again has to do with normalization to a flux of incoming particles. The slashed notation is in homage to Hawking's infamous ``dollar matrix''~\cite{hawking1976breakdown}. The $\slashed{M}$ elements can be computed by a simple set of Feynman rules, which we depict schematically in Fig. \ref{fig:scattering}.

A remarkable result in our explicit calculations is that the scattering supermatrix elements defined by Eq. \eqref{eq:slashM} transform covariantly under the Poincar\'e group. We give some more precise statements to this effect below and in Appendix \ref{app:lorentz}. In particular, note that we are computing transition elements only between states of the quantum matter, with the $\phi$ field averaged over. It is not at all obvious that this result should be covariant --- for example, one might have imagined that the classical field evolved into some state which spontaneously broke Lorentz invariance by setting a direction. As a sanity check, we show in Appendix \ref{app:lorentz} that an analogous statement holds if we treat $\phi$ as a quantum field, initialize it to the vacuum $\ket{0}$, and trace over the final states. Even allowing for processes where $\phi$ quanta are emitted into the final state, we find that the trace produces a channel on the $\boldsymbol{\chi}$ fields which is Lorentz covariant. The key seems to be that the initial state is Lorentz invariant.

One major conceptual difference between CQ theory and ordinary quantum scattering is that the asymptotic states in CQ theory are necessarily unstable. There are two reasons for this. One is that there is no mass gap for the classical field, even if it has a non-zero mass, because one can always find solutions to the wave equation with arbitrarily small energy. The other is that the classical field is being continuously stochastically driven. In particular, this means that the initial state we will use, $\mathcal{P}[\phi,\pi] = \delta[\phi] \delta[\pi]$, is completely unstable. In fact, at sufficiently long times, any initial classical field configuration will asymptote to $\mathcal{P}[\phi,\pi] \equiv 1$, because the evolution law is purely diffusive with no friction. This is a highly pathological state, especially once extrapolated to gravity. It would represent a state where the universe is equally likely to be in flat spacetime or a highly chaotic assortment of gravitational excitations. In particular, the CQ model does not have a vacuum state for the classical fields. Quite similarly, momentum eigenstates for the quantum particles are also unstable, because the $D_0$ term is trying to decohere them into the field configuration basis.

One could attempt to deal with these problems by adiabatically switching off the $D_{0,2}$ terms. However, perhaps a more straightforward interpretation is to recognize that a scattering description in this model only makes sense for finite times $T$ such that the $D_{0,2}$ effects are small. In particular, we are going to treat this model in perturbation theory, so $T D_0$ and $T D_2$ cannot be too large anyway. Thus in what follows we will regulate everything with an overall time duration $T < \infty$. This is similar to the use of scattering theory to compute transitions and decay rates with quasistable particles in quantum field theory, for example muon scattering.

With these assumptions and caveats in mind, we begin by calculating $2 \to 2$ scattering of the $\chi$ particles in the non-relativistic limit. This is an instructive warmup to show how standard path integral scattering calculations generalize to this setting. We then move on to a relativistic calculation to analyze the Lorentz invariance of the scattering process.

\subsection{Non-relativistic $2 \to 2$ Yukawa scattering}
\label{sbsc:NR}

To begin, we will analyze $\chi_1 \chi_2 \to \chi_1 \chi_2$ scattering by starting directly in the non-relativistic limit for the massive particles. That is, we assume that $\mb{p}_i/m_i \ll 1$ and work with single-particle states only. We continue to use a continuum description of the Yukawa field $\phi$. The interaction takes the form
\be
V_{\rm int} = \frac{\lambda}{m} \sum_{i=1,2} \int d^3\mb{x} \, \phi(\mb{x}) \delta^3(\mb{x} - \mb{x}_i),
\ee
using standard matching methods. The free Lagrangian for the particles is $L_0[\mb{x}_1,\mb{x}_2] = m \sum_i \dot{\mb{x}}^2_i/2$. Our main goal here is to show how one can apply elementary scattering methods to the CQ framework. We will extend the beautiful treatment of non-relativistic quantum scattering given in~\cite{ryder1996quantum}.

As discussed in Sec. \ref{sec:reduced}, the reduced dynamics of the particles can be calculated by first performing the path integral integral over $\phi$. This produces a Feynman-Vernon type path integral for the quantum particles,
\begin{align}
\begin{split}
\label{fv-nr}
\rho(\mb{x}_{1,f}, \mb{x}_{2,f}, \ul{\mb{x}}_{1,f}, \ul{\mb{x}}_{2,f}, t_f) & = \int D\mb{x}_1 D\mb{x}_2 D\ul{\mb{x}}_1 D\ul{\mb{x}}_2 e^{i (S_0[\mb{x}_1,\mb{x}_2] - S_0[\ul{\mb{x}}_1,\ul{\mb{x}}_2])} e^{i S_{IF}[\mb{x}_1,\mb{x}_2,\ul{\mb{x}}_1,\ul{\mb{x}}_2]} \\
& \times \rho(\mb{x}_{1,0}, \mb{x}_{2,0}, \ul{\mb{x}}_{1,0}, \ul{\mb{x}}_{2,0} t_0).
\end{split}
\end{align}
In this simple example, we have
\begin{align}
\label{IF-nr}
i S_{IF} = -\frac{\lambda^2}{m_1 m_2} \int dt \, F_1(\mb{x}_1-\mb{x}_2) + F_1^*(\ul{\mb{x}}_1-\ul{\mb{x}}_2) - \sum_{i\neq j} F_2(\mb{x}_i - \ul{\mb{x}}_j)
\end{align}
and the boundary values of all the integrals are taken at $t_f$ and $t_0$, which will later become $\pm T/2$. The functions here are
\begin{align}
\begin{split}
\label{f-and-g}
F_1(\mb{x}_1 - \mb{x}_2) & = - \frac{i}{2} \left[ G_R(\mb{x}_1 - \mb{x}_2) + G_R(\mb{x}_2 - \mb{x}_1) \right] + D_2 G_C(\mb{x}_1 - \mb{x}_2) + D_0 \delta^3(\mb{x}_1 - \mb{x}_2) \\
F_2(\mb{x}_1 - \mb{x}_2) & =  D_2 G_C(\mb{x}_1 - \mb{x}_2) + D_0 \delta^3(\mb{x}_1 - \mb{x}_2)
\end{split}
\end{align}
with
\be
G_R(\mb{x}) = \int \frac{d^3\mb{k}}{(2\pi)^3} \frac{e^{i \mb{k} \cdot \mb{x}}}{\mb{k}^2 + m_{\phi}^2}, \ \ G_C(\mb{x}) = \int \frac{d^3\mb{k}}{(2\pi)^3} \frac{e^{i \mb{k} \cdot \mb{x}}}{(\mb{k}^2 + m_{\phi}^2)^2}, \ \ \delta^3(\mb{x})  = \int \frac{d^3\mb{k}}{(2\pi)^3} e^{i \mb{k} \cdot \mb{x}}.
\ee
The subscript $R$ on the first Green's function stands for retarded; this notation is just an artifact of the full relativistic calculation shown in the next section. The subscript $C$ on the second Green's function stands for classical; it comes from the Fokker-Planck path integral over $\phi$, which is quartic in derivatives as discussed above. 

The expression in Eq. \eqref{IF-nr} is a highly simplified form of the fully relativistic expression. We give a detailed derivation of $S_{IF}$ in Appendix \ref{app:fv}. To go from the relativistic case to Eq. \eqref{IF-nr}, we are assuming that all momenta are non-relativistic, and dropping all self-interaction terms. This latter assumption is particularly helpful, although to justify it would require a detailed study of renormalization of propagators in this theory, which goes beyond the scope of what we are trying to do here. We make some further comments in Appendix \ref{app:fv}.

\begin{figure}[t]

\begin{tikzpicture}[scale=.43]


\draw [->] (-4,-3) -- (-4,3);
\draw [->] (-2,-3) -- (-2,3);

\node at (-4,-4) {$\mb{p}_1$};
\node at (-2,-4) {$\mb{p}_2$};

\node at (-4,4) {$\mb{p}_1'$};
\node at (-2,4) {$\mb{p}_2'$};

\draw [<-] (2,-3) -- (2,3);
\draw [<-] (4,-3) -- (4,3);

\node at (2,-4) {$\ul{\mb{p}}_1$};
\node at (4,-4) {$\ul{\mb{p}}_2$};

\node at (2,4) {$\ul{\mb{p}}_1'$};
\node at (4,4) {$\ul{\mb{p}}_2'$};

\draw [fill=lightgray] (0,0) ellipse [x radius=4.5, y radius=1];

\node at (6,0) {$=$};


\begin{scope}[shift={(10,0)}]

\node [yscale=8,xscale=2] at (-3,0) {(};

\draw [->] (-2,-3) -- (-2,3);
\draw [->] (-1,-3) -- (-1,3);

\node at (0,0) {$\times$};

\draw [<-] (1,-3) -- (1,3);
\draw [<-] (2,-3) -- (2,3);

\node [yscale=8,xscale=2] at (3,0) {)};

\node at (4,0) {$+$};

\end{scope}


\begin{scope}[shift={(18,0)}]

\node [yscale=8,xscale=2] at (-3,0) {(};

\draw [->] (-2,-3) -- (-2,3);
\draw [->] (-1,-3) -- (-1,3);
\draw [thick,dotted] (-2,0) -- (-1,0);

\node at (0,0) {$\times$};

\draw [<-] (1,-3) -- (1,3);
\draw [<-] (2,-3) -- (2,3);

\node [yscale=8,xscale=2] at (3,0) {)};

\node at (4,0) {$+$};

\end{scope}


\begin{scope}[shift={(26,0)}]

\node [yscale=8,xscale=2] at (-3,0) {(};

\draw [->] (-2,-3) -- (-2,3);
\draw [->] (-1,-3) -- (-1,3);

\draw [thick,dotted] (-1,0) -- (1,0);

\draw [<-] (1,-3) -- (1,3);
\draw [<-] (2,-3) -- (2,3);

\node [yscale=8,xscale=2] at (3,0) {)};

\node at (4,0) {$+$};
\node at (5.5,0) {$\cdots$};

\end{scope}

\end{tikzpicture}

\caption{Schwinger-Keldysh-Feynman-Vernon diagrams for $2 \to 2$ scattering in CQ Yukawa theory, in the non-relativistic limit, to $\Oi(\lambda^2)$. The dotted line indicates interaction through the Yukawa field. The first term on the right hand side represents no interaction at all, i.e., the $\lambda^0$ term. The second term gives an example of ket-ket interaction ($\sim M_{F_1}$). The third gives an example gives an example of ket-bra interaction ($\sim M_{F_2}$). The dots represent the other permutations and higher order terms.}
\label{fig:scattering-2}

\end{figure}
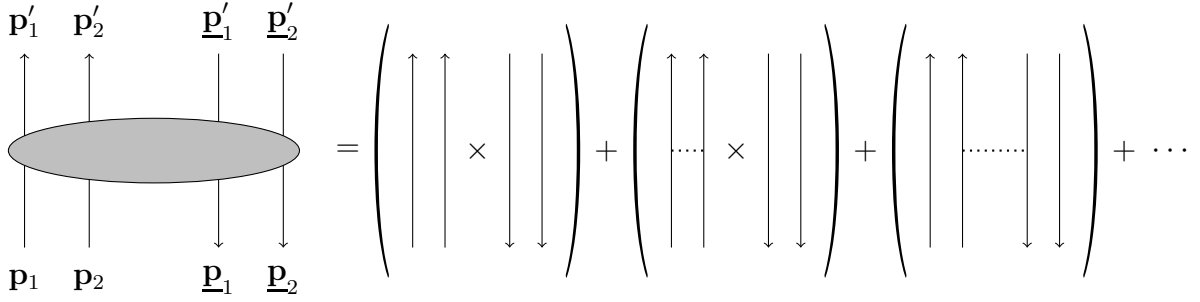

While the full relativistic Feynman-Vernon functional is non-local in time, our non-relativistic limit Eq. \eqref{fv-nr} is local in time. Thus it can be used to do scattering calculations in nearly the exact same fashion as ordinary non-relativistic quantum mechanics. The only new things are that the ``potential'' defined by $i S_{IF} = \int dt\, V_{IF}(t)$ is complex and it can connect bra and ket variables.  

Consider $2 \to 2$ scattering. We let $\mb{p}_i, \ul{\mb{p}}_i$ denote the initial momenta in the ket and bra respectively, and put primes for final state momenta. The full scattering supermatrix element is computed by evolving an initial set of plane wave states
\be
\rho(\mb{x}_1,\mb{x}_2,\ul{\mb{x}}_1,\ul{\mb{x}}_2,t_0) = \prod_{i,j=1,2} \psi_{\mb{p}_i}(\mb{x}_{i,0},t_0) \psi^*_{\ul{\mb{p}}_j}(\ul{\mb{x}}_{j,0},t_0), \ \ \ \psi_{\mb{p}}(\mb{x},t) = \frac{e^{i (\mb{p} \cdot \mb{x} - E_{\mb{p}} t)}}{(2\pi)^{3/2}} 
\ee
and then projecting the result onto a similar set of plane waves at $t_f$. Concretely,
\begin{align}
\begin{split}
\label{eq:slashedM-NR}
\slashed{M}_{\mb{p}_1\mb{p}_2\ul{\mb{p}}_1\ul{\mb{p}}_2 \to \mb{p}_1' \mb{p}_2' \ul{\mb{p}}_1' \ul{\mb{p}}_2'} & =  \int  d\mb{x}_{1,f} d\mb{x}_{2,f} d\ul{\mb{x}}_{1,f} d\ul{\mb{x}}_{2,f} d\mb{x}_{1,0} d\mb{x}_{2,0} d\ul{\mb{x}}_{1,0} d\ul{\mb{x}}_{2,0} \\
& \times 
\psi^*_{\mb{p}'_1}(\mb{x}_{1,f},t_f) 
\psi^*_{\mb{p}'_2}(\mb{x}_{2,f},t_f)
\psi_{\ul{\mb{p}}'_1}(\ul{\mb{x}}_{1,f},t_f) \psi_{\ul{\mb{p}}'_2}(\ul{\mb{x}}_{2,f},t_f)
\\
& \times \slashed{K}(\mb{x}_{1,f}, \mb{x}_{2,f}, \ul{\mb{x}}_{1,f}, \ul{\mb{x}}_{2,f}, t_f | \mb{x}_{1,0}, \mb{x}_{2,0}, \ul{\mb{x}}_{1,0}, \ul{\mb{x}}_{2,0}, t_0) \\
& \times 
\psi_{\mb{p}_1}(\mb{x}_{1,0},t_0) 
\psi_{\mb{p}_2}(\mb{x}_{2,0},t_0)
\psi^*_{\ul{\mb{p}}_1}(\ul{\mb{x}}_{1,0},t_0) \psi^*_{\ul{\mb{p}}_2}(\ul{\mb{x}}_{2,0},t_0).
\end{split}
\end{align}
Here, the density matrix propagator $\slashed{K}$ from $t_0$ to $t_f$ is given by
\begin{align}
\begin{split}
\label{K-NR}
& \slashed{K}(\mb{x}_{1,f}, \mb{x}_{2,f}, \ul{\mb{x}}_{1,f}, \ul{\mb{x}}_{2,f}, t_f | \mb{x}_{1,0}, \mb{x}_{2,0}, \ul{\mb{x}}_{1,0}, \ul{\mb{x}}_{2,0}, t_0) \\
& \ \ \ \ = \int D\mb{x}_1 D\mb{x}_2 D\ul{\mb{x}}_1 D\ul{\mb{x}}_2 \, e^{i (S_0[\mb{x}_1,\mb{x}_2] - S_0[\ul{\mb{x}}_1,\ul{\mb{x}}_2])} e^{i S_{IF}[\mb{x}_1,\mb{x}_2,\ul{\mb{x}}_1,\ul{\mb{x}}_2]},
\end{split}
\end{align}
with the limits on the path integral given by the argument of $\slashed{K}$. To compute this perturbatively, we expand $e^{i S_{IF}}$ as a power series in $\lambda/m$, which gives an expression of the schematic form
\be
\slashed{K}(t_f|t_0) = \slashed{K}_0(t_f|t_0) + \int dt \, \slashed{K}_0(t_f|t) V_{IF}(t) \slashed{K}_0(t|t_0) + \cdots
\ee
where $\slashed{K}_0$ is the free propagator.

The free propagator is standard: it is just a product over each particle,
\begin{align}
\slashed{K}_0(\mb{x}_{1,f}, \mb{x}_{2,f}, \ul{\mb{x}}_{1,f}, \ul{\mb{x}}_{2,f}, t_f | \mb{x}_{1,0}, \mb{x}_{2,0}, \ul{\mb{x}}_{1,0}, \ul{\mb{x}}_{2,0}, t_0) =  \prod_{i,j} K_0(\mb{x}_{i,f},t_f | \mb{x}_{i,0},t_0) K_0^*(\ul{\mb{x}}_{j,f},t_f | \ul{\mb{x}}_{j,0},t_0), 
\end{align}
where $K_0$ is the ordinary propagator for a single non-relativistic particle in quantum mechanics (see~\cite{ryder1996quantum} for the explicit expression). The term $\slashed{K}_2$ of order $\lambda^2/(m_1 m_2)$ is more complicated:
\begin{align}
\begin{split}
\slashed{K}_2 & =  -\frac{\lambda^2}{m_1 m_2} \int dt d\mb{y}_1 d\mb{y}_2 d\ul{\mb{y}}_1 d\ul{\mb{y}}_2 \, \slashed{K}_0(\mb{x}_{1,f}, \mb{x}_{2,f}, \ul{\mb{x}}_{1,f}, \ul{\mb{x}}_{2,f}, t_f | \mb{y}_{1}, \mb{y}_{2}, \ul{\mb{y}}_{1}, \ul{\mb{y}}_{2}, t) \\
& \times  \left[ F_1(\mb{y}_1-\mb{y}_2) + F_1^*(\ul{\mb{y}}_1-\ul{\mb{y}}_2) - \sum_{i \neq j} F_2(\mb{y}_i - \ul{\mb{y}}_j) \right] \\
& \times  \slashed{K}_0(\mb{y}_{1}, \mb{y}_{2}, \ul{\mb{y}}_{1}, \ul{\mb{y}}_{2}, t | \mb{x}_{1,0}, \mb{x}_{2,0}, \ul{\mb{x}}_{1,0}, \ul{\mb{x}}_{2,0}, t_0 ).
\end{split}
\end{align}
Here the $\mb{y}$ are the variables at the time $t$ where the interaction occurs. While this equation is long, it is easy to understand in terms of Feynman-type diagrams. See Fig. \ref{fig:scattering-2}. To leading order ($\Oi(\lambda^2)$) in perturbation theory, we have a simple factorization of the supermatrix elements:
\begin{align}
\label{eq:slashedM}
\begin{split}
& \slashed{M}_{\mb{p}_1\mb{p}_2\ul{\mb{p}}_1\ul{\mb{p}}_2 \to \mb{p}_1' \mb{p}_2' \ul{\mb{p}}_1' \ul{\mb{p}}_2'} = \delta^3(\mb{p}_1 - \mb{p}_1') \delta^3(\mb{p}_2 - \mb{p}_2') \delta^3(\ul{\mb{p}}_1 - \ul{\mb{p}}_1') \delta^3(\ul{\mb{p}}_2 - \ul{\mb{p}}_2') \\
& + M_{F_1}(\mb{p}_1 \mb{p}_2 \to \mb{p}_1' \mb{p}_2') \delta^3(\ul{\mb{p}}_1 - \ul{\mb{p}}_1') \delta^3(\ul{\mb{p}}_2 - \ul{\mb{p}}_2') + \delta^3(\mb{p}_1 - \mb{p}_1') \delta^3(\mb{p}_2 - \mb{p}_2') M_{F_1^*}(\ul{\mb{p}}_1 \ul{\mb{p}}_2 \to \ul{\mb{p}}_1' \ul{\mb{p}}_2')  \\
& + \delta^3(\ul{\mb{p}}_2 - \ul{\mb{p}}_2') M_{F_2}(\mb{p}_1 \ul{\mb{p}}_2 \to \mb{p}_1' \ul{\mb{p}}_2')  \delta^3(\ul{\mb{p}}_1 - \ul{\mb{p}}_1') + \delta^3(\ul{\mb{p}}_1 - \ul{\mb{p}}_1') M_{F_2}(\mb{p}_2 \ul{\mb{p}}_1 \to \mb{p}_2' \ul{\mb{p}}_1')  \delta^3(\ul{\mb{p}}_2 - \ul{\mb{p}}_2').
\end{split}
\end{align}
The term in the first line represents no interaction at all. The following terms contain the effects of either the $F_1$ interaction which connects ket-ket or bra-bra, and the $F_2$ interaction which connects kets to bras. We now turn to explicit calculation of these terms.

First consider the term with $F_1$. This clearly factors in the sense that it affects only the ket, not the bra. Inserting it into Eq. \eqref{eq:slashedM-NR}, we have a contribution to the scattering supermatrix
\begin{align}
\begin{split}
& M_{F_1}(\mb{p}_1\mb{p}_2 \to \mb{p}_1'\mb{p}_2') =-\frac{\lambda^2}{m_1 m_2}  \int  dt d\mb{y}_1 d\mb{y}_2 d\mb{x}_{1,f} d\mb{x}_{2,f} d\mb{x}_{1,0} d\mb{x}_{2,0}
\psi^*_{\mb{p}'_1}(\mb{x}_{1,f},t_f) 
\psi^*_{\mb{p}'_2}(\mb{x}_{2,f},t_f)
\\
& \times  K_0(\mb{x}_{1,f}, \mb{x}_{2,f}, t_f | \mb{y}_{1}, \mb{y}_{2}, t) F_1(\mb{y}_1-\mb{y}_2) K_0(\mb{y}_{1}, \mb{y}_{2}, t | \mb{x}_{1,0}, \mb{x}_{2,0}, t_0) \psi_{\mb{p}_1}(\mb{x}_{1,0},t_0) \psi_{\mb{p}_2}(\mb{x}_{2,0},t_0).
 \end{split}
 \end{align}
Performing the initial and final state integrals moves the exponentials to the interaction vertex at $t$,
\begin{align}
\begin{split}
\label{MF1-simple}
M_{F_1}(\mb{p}_1\mb{p}_2 \to \mb{p}_1'\mb{p}_2') & =-\frac{\lambda^2}{m_1 m_2}  \int  dt d\mb{y}_1 d\mb{y}_2 F_1(\mb{y}_1-\mb{y}_2) 
\psi^*_{\mb{p}'_1}(\mb{y}_1,t) 
\psi^*_{\mb{p}'_2}(\mb{y}_2,t)
\psi_{\mb{p}_1}(\mb{y}_1,t) 
\psi_{\mb{p}_2}(\mb{y}_2,t),
\end{split}
\end{align}
where the free propagator gave the free evolution of the plane waves,
\begin{align}
\begin{split}
\psi_{\mb{p}_1}(\mb{y}_{1},t) 
\psi_{\mb{p}_2}(\mb{y}_{2},t)
&=\int  d\mb{x}_{1,0} d\mb{x}_{2,0} \, K_0(\mb{y}_{1}, \mb{y}_{2}, t | \mb{x}_{1,0}, \mb{x}_{2,0}, t_0)
\psi_{\mb{p}_1}(\mb{x}_{1,0},t_0) 
\psi_{\mb{p}_2}(\mb{x}_{2,0},t_0)
\\
\psi^*_{\mb{p}'_1}(\mb{y}_{1},t) 
\psi^*_{\mb{p}'_2}(\mb{y}_{2},t)
&=\int d\mb{x}_{1,f} d\mb{x}_{2,f} 
\, \psi^*_{\mb{p}'_1}(\mb{x}_{1,f},t_f) 
\psi^*_{\mb{p}'_2}(\mb{x}_{2,f},t_f)
K_0(\mb{x}_{1,f},\mb{x}_{2,f}, t_f |\mb{y}_{1}, \mb{y}_{2} , t).
\end{split}
\end{align}
In Eq. \eqref{MF1-simple}, the integral over $dt$ produces $(2\pi) \delta(E_{\mb{p}_1'} + E_{\mb{p}_2'} - E_{\mb{p}_1} - E_{\mb{p}_2})$, i.e., total energy conservation. Now we insert the expressions in Eq. \eqref{f-and-g} for $F_1$. Every term in $F_1$ can be expressed as an integral over $d^3\mb{k}$ with a factor $e^{\pm i \mb{k} \cdot (\mb{y}_1 - \mb{y}_2)}$. The integrals over the $\mb{y}$ variables then produce a pair of momentum delta functions $(2\pi)^6 \delta^3(\mb{p}_1' - \mb{p}_1 \pm \mb{k}) \delta^3(\mb{p}_2' - \mb{p}_2 \mp \mb{k})$. Writing this as an overall momentum conservation delta function, the $\mb{k}$ integral can now be performed by setting $\mb{k} = \mb{p}_1' - \mb{p}_1$. This produces the simple answer
\begin{align}
\label{MF1-final}
M_{F_1}(\mb{p}_1\mb{p}_2 \to \mb{p}_1'\mb{p}_2')  = -\delta^4(p_1 + p_2 - p_1' - p_2') \frac{\lambda^2}{(2\pi)^2 m_1 m_2} \left[ \frac{i}{\mb{k}^2 + m_{\phi}^2} + \frac{D_2}{(\mb{k}^2 + m_{\phi}^2)^2} +  D_0 \right]_{\mb{k} = \mb{p}_1' - \mb{p}_1}.
\end{align}
The first term is ordinary scattering in quantum mechanics. The next two terms are open system corrections and take a specific form in CQ theory. We have written the total energy-momentum delta function in a Lorentz-invariant notation, but it should be remembered that in this limit the energies do not actually transform correctly to make four-vectors. The term with $F_1^*$ is clearly the same except we replace the ket variables with bra variables. 

Now we analyze the $F_2$ terms, which mix the bra and ket variables. This no longer factorizes between the bra and ket. However, one particle in the ket and one in the bra will still evolve trivially. Using the same manipulations as above that led to Eq. \eqref{MF1-simple}, we obtain
\begin{align}
\begin{split}
M_{F_2}(\mb{p}_i\ul{\mb{p}}_j \to \mb{p}_i' \ul{\mb{p}}_j') & = - \frac{\lambda^2}{m_1 m_2}  \int  dt d\mb{y}_i d\ul{\mb{y}}_j F_2(\mb{y}_i-\ul{\mb{y}}_j) 
\psi^*_{\mb{p}'_i}(\mb{y}_i,t) 
\psi_{\ul{\mb{p}}'_j}(\ul{\mb{y}}_j,t)
\psi_{\mb{p}_i}(\mb{y}_i,t) 
\psi^*_{\ul{\mb{p}}_j}(\ul{\mb{y}}_j,t).
\end{split}
\end{align}
There is a critical difference between this and Eq. \eqref{MF1-simple}: the phases on the bra variables are flipped. When we perform the integral over the vertex variables, this now produces a different set of delta functions: a momentum conservation factor $\sim \delta^3(\mb{p}_i' - \mb{p}_i - (\ul{\mb{p}}_j' - \ul{\mb{p}}_j))$ and an energy conservation factor $\sim \delta(E_{\mb{p}_i'} - E_{\mb{p}_i}  - ( E_{\ul{\mb{p}}_j'} - E_{\ul{\mb{p}}_j}))$. In total, we get
\begin{align}
\label{MF2}
M_{F_2}(\mb{p}_i \ul{\mb{p}}_j \to \mb{p}_i' \ul{\mb{p}}_j')  = - \delta^4(p_i' - p_i - (\ul{p}_j'  - \ul{p}_j)) \frac{\lambda^2}{(2\pi)^2 m_1 m_2} \left[\frac{D_2}{(\mb{k}^2 + m_{\phi}^2)^2} +  D_0 \right]_{\mb{k} = \mb{p}_i' - \mb{p}_i}.
\end{align}
We have written the arguments of the delta functions in a suggestive order: $M_{F_2}$ represents a process where the scalar $\phi$ imparts an equal kick $\mb{k} = \Delta \mb{p}$ to both the ket and the bra particle. 

Finally, we can use these results to calculate the actual transition probabilities, using Eqs.~\eqref{eq:Pgeneral} and \eqref{eq:slashedM}. To do this we are going to look at diagonal elements $\mb{p}_i = \ul{\mb{p}}_i$ and $\mb{p}_i' = \ul{\mb{p}}_i'$. To order $\lambda^2$, the only non-trivial process is forward scattering $\mb{p}_i = \mb{p}_i'$; anything else vanishes due to the various delta functions. As a non-trivial check on the CQ framework, we show in Appendix \ref{app-probability} that these forward scattering terms sum appropriately to conserve total probability (i.e., the open systems version of the optical theorem holds). However, to get a differential cross section for non-trivial scattering angles, we are therefore going to need to go to order $\lambda^4$. 

The $\lambda^2$ terms in $M_{F_1}$ and $M_{F_1^*}$ multiply to produce non-trivial scattering at this order, namely
\be
\slashed{M}_{\mb{p}_1 \mb{p}_2 \mb{p}_1 \mb{p}_2 \to \mb{p}'_1 \mb{p}'_2 \mb{p}'_1 \mb{p}'_2} \supset \left| M_{F_1}(\mb{p}_1 \mb{p}_2 \to \mb{p}_1' \mb{p}_2') \right|^2,
\ee
with $M_{F_1}$ given in Eq. \eqref{MF1-final}. Other than the CQ corrections, this works like ordinary scattering. There is also a contribution to forward scattering at this order, where the evolution acts to order $\lambda^4$ but only on the ket, or only on the bra --- i.e., the one-loop forward scattering contribution, which we will ignore following the above discussion.

What about the $F_2$ terms? To order $\lambda^4$, we can have two kinds of contributions. One is the process where $\ket{\mb{p}_1}$ interacts with $\bra{\mb{p}_2}$ and $\ket{\mb{p}_2}$ interacts with $\bra{\mb{p}_1}$, which is a product of two tree diagrams. The other possibility is a loop diagram, where e.g. $\ket{\mb{p}_1}$ interacts with $\bra{\mb{p}_2}$ twice. It is easy to check that the loop diagrams only contribute to forward scattering again to this order, so we will continue to ignore them. Thus the $F_2$ term contributes as
\begin{align}
\slashed{M}_{\mb{p}_1 \mb{p}_2 \mb{p}_1 \mb{p}_2 \to \mb{p}'_1 \mb{p}'_2 \mb{p}'_1 \mb{p}'_2}\supset M_{F_2}(\mb{p}_{1} \mb{p}_{2} \to \mb{p}_{1}' \mb{p}_{2}') M_{F_2}(\mb{p}_{2} \mb{p}_{1} \to \mb{p}_{2}' \mb{p}_{1}') =  \left[ M_{F_2}(\mb{p}_{1} \mb{p}_{2} \to \mb{p}_{1}' \mb{p}_{2}') \right]^2,
\end{align}
where in the last equality we used Eq. \eqref{MF2}. 

Putting the above results together we obtain the total probability. Both the $F_1$ and $F_2$ contributions contain squares of energy-momentum conservation delta functions. To regulate this, we can do the usual process of converting one of the delta functions into a spacetime volume $VT/(2\pi)^4$~\cite{weinberg2015lectures,weinberg1995quantum}. Using this, we have, for non-forward scattering, 
\begin{align}
\begin{split}
\label{eq:P-NR}
& P(\mb{p}_1 \mb{p}_2 \to \mb{p}_1' \mb{p}_2') = \left| M_{F_1}(\mb{p}_1\mb{p}_2 \to \mb{p}_1' \mb{p}_2') \right|^2 + \left[ M_{F_2}(\mb{p}_{1} \mb{p}_{2} \to \mb{p}_{1}' \mb{p}_{2}') \right]^2 \\
& = \frac{VT}{(2\pi)^4} \delta^4(p_1' + p_2' - p_1 - p_2) \frac{\lambda^4}{(2\pi)^4 m^2_1 m^2_2} \left[ \left( \frac{1}{(\mb{p}_1' - \mb{p}_1)^2 + m_{\phi}^2} \right)^2 + \left( \frac{D_2}{[(\mb{p}_1' - \mb{p}_1)^2 + m_{\phi}^2]^2} +  D_0 \right)^2 \right] \\
& + \frac{VT}{(2\pi)^4} \delta^4(p_1' - p_1 - (p_2' - p_2)) \frac{\lambda^4}{(2\pi)^4 m^2_1 m^2_2} \left(   \frac{D_2}{[(\mb{p}_1' - \mb{p}_1)^2 + m_{\phi}^2]^2} +  D_0 \right)^2.
\end{split}
\end{align}
We see that the two momentum transfers are the same $\mb{p}_1'-\mb{p}_1$, but the energy-momentum delta functions are different. In particular, this is a Galilean-invariant expression, but includes non-trivial probabilities for processes which violate momentum conservation. As mentioned above, this is consistent because this is calculated in an open quantum system, and so there is no Noether theorem connecting conservation laws to symmetries. Note that large violations of momentum conservation are suppressed like $1/\Delta p^4$.

Finally, we can derive a differential cross section from Eq. \eqref{eq:P-NR} following standard methods~\cite{weinberg2015lectures}. Let $u$ be the relative velocity of the initial state momenta, e.g., $u = |\mb{p}|/\mu$ in the center-of-momentum frame, where $\mu$ is the reduced mass. Then the differential rate of particles scattered into the final states in a phase space volume $d^3\mb{p}_1' d^3\mb{p}_2'$ is given by
\begin{align}
\begin{split}
\label{dsigma-nr}
\frac{d\sigma}{d^3\mb{p}_1' d^3\mb{p}_2'} & = \frac{(2\pi)^2}{u} \frac{\lambda^4}{(2\pi)^4 m_1^2 m_2^2} \\
& \times \Bigg\{ \delta^4(p_1' + p_2' - p_1 - p_2) \left[ \left( \frac{1}{(\mb{p}_1' - \mb{p}_1)^2 + m_{\phi}^2} \right)^2 + \left( \frac{D_2}{[(\mb{p}_1' - \mb{p}_1)^2 + m_{\phi}^2]^2} +  D_0 \right)^2 \right] \\
& + \delta^4(p_1' - p_1 - (p_2' - p_2)) \left(   \frac{D_2}{[(\mb{p}_1' - \mb{p}_1)^2 + m_{\phi}^2]^2} +  D_0 \right)^2 \Bigg\}.
\end{split}
\end{align}
In Sec. \ref{sec:example}, we show how to evaluate this in a frame in an explicit example, the scattering of a light particle off a heavy one. 

The non-relativistic example here demonstrates the basic structure of scattering in CQ theory. It is also sufficient to extract useful predictions, as discussed in Sec. \ref{sec:example}. However, it is of considerable formal interest to understand the full relativistic version of this calculation, to which we now turn.

\subsection{Relativistic $2 \to 2$ Yukawa scattering}
\label{subsec:rscattering}

Let us move on to the analysis of $\chi_1 \chi_2 \rightarrow \chi_1 \chi_2$ scattering without taking the non-relativistic limit. We use the relativistic interaction
of Eq.~\eqref{eq:Yukawa} and consider the massive scalar fields 
$\boldsymbol{\chi}=(\chi_1, \chi_2)$ in ordinary canonical quantization. We find the following Feynman-Vernon type path integral for the fields, 
\begin{align}
\begin{split}
\label{fv-rel}
\rho(\chi_{1,f}, \chi_{2,f}, \ul{\chi}_{1,f}, \ul{\chi}_{2,f}, t_f) & = \int D\chi_1 D\chi_2  D\ul{\chi}_1 D\ul{\chi}_2 e^{i (S_0[\boldsymbol{\chi}] - S_0 [\ul{\boldsymbol{\chi}}])} e^{i S_{IF}[\boldsymbol{\chi},\ul{\boldsymbol{\chi}}]}  \rho(\chi_{1,0}, \chi_{2,0}, \ul{\chi}_{1,0}, \ul{\chi}_{2,0}, t_0),
\end{split}
\end{align}
where $S_0 [\boldsymbol{\chi}]=-\frac{1}{2} \int d^4x \sum_i [(\partial_\mu \chi_i)^2+m^2_i \chi^2_i]$ is the free action of the fields and 
\be
\label{IF-r}
i S_{IF} = -\lambda^2  \int d^4x d^4y \, 
F_1(x-y)\chi^2_1 (x) \chi^2_2 (y)+F^*_1(x-y) \ul{\chi}^2_1 (x) \ul{\chi}^2_2 (y) - \sum_{i\neq j} F_2(x-y) \chi^2_i (x) \ul{\chi}^2_j (y). 
\ee
is the influence functional. Again see Appendix \ref{app:fv} for details. Note that unlike the non-relativistic case, this is a non-local functional. Here, we again only wrote the cross terms between the two fields, dropping the self-interaction terms. These only contribute to loops in the forward scattering process and are beyond the scope of this paper. The $F_{1,2}$ functions here are 
\begin{align}
F_1 (x-y)&=-\frac{i}{2} [G_R (x-y)+G_R (y-x)]+D_2 G_C (x-y)+D_0 \delta^4(x-y), 
\\
F_2 (x-y)&= 
\frac{i}{2} [G_R (x-y)-G_R (y-x)]+D_2 G_C (x-y)+D_0 \delta^4(x-y)
\end{align}
with the relativistic forms 
\begin{align}
\begin{split}
G_R(x) &= \int \frac{d^4k}{(2\pi)^4} \frac{e^{i k \cdot x}}{-(k^0+i\epsilon)^2 +\mb{k}^2 + m_{\phi}^2}, \\ 
G_C(x) &= \int \frac{d^4k}{(2\pi)^4} \frac{e^{i k \cdot x}}{[-(k^0+i\epsilon)^2+\mb{k}^2 + m_{\phi}^2][-(k^0-i\epsilon)^2+\mb{k}^2 + m_{\phi}^2]}, 
\\ 
\delta^4(x) &= \int \frac{d^4k}{(2\pi)^4} e^{i k\cdot x}.
\end{split}
\end{align}
The retarded Green's function describes the exchange of energy and momentum between 
$\chi_1$ and 
$\chi_2$.
One can reproduce the non-relativistic counterpart \eqref{IF-nr} by taking 
$\chi^2_i (x) \rightarrow \frac{1}{m} \delta^3(\mb{x}-\mb{x}_i)$ and ignoring 
$k^0 \pm i\epsilon $ in the functions $G_R$ and $G_C$. 

In this relativistic setting, the scattering supermatrix element is given by the path integral,
\begin{align}
\begin{split}
\label{eq:slashedM-R}
\slashed{M}_{p_1 p_2\ul{p}_1\ul{p}_2 \to p_1' p_2' \ul{p}_1' \ul{p}_2'}
& 
=\int D\chi_{1,f} D \chi_{2,f}D\ul{\chi}_{1,f} D \ul{\chi}_{2,f} D\chi_{1,0} D \chi_{2,0} D\ul{\chi}_{1,0} D \ul{\chi}_{2,0} 
\\
&
\times 
\Psi^*_{p'_1} (\chi_{1,f},t_f) 
\Psi^*_{p'_2} (\chi_{2,f},t_f) 
\Psi_{\ul{p}'_1} (\ul{\chi}_{1,f},t_f) 
\Psi_{\ul{p}'_2} (\ul{\chi}_{2,f},t_f) 
\\
&
\times 
\slashed{K}(\chi_{1,f},\chi_{2,f},\ul{\chi}_{1,f},\ul{\chi}_{2,f},t_f|\chi_{1,0},\chi_{2,0},\ul{\chi}_{1,0},\ul{\chi}_{2,0},t_0)
\\
&
\times 
\Psi_{p_1}(\chi_{1,0},t_0) 
\Psi_{p_2}(\chi_{2,0},t_0) 
\Psi^*_{\ul{p}_1}(\ul{\chi}_{1,0},t_0) 
\Psi^*_{\ul{p}_2}(\ul{\chi}_{2,0},t_0),
\end{split}
\end{align}
where 
$\Psi_{p_i} (\chi_i,t)=e^{-i\omega_{\mb{p}_i} t} \langle \chi_i |\mb{p}_i \rangle$ with 
$\omega_{\mb{p}_i}=\sqrt{\mb{p}^2_i+m_i^2}$ is the wave functional of momentum eigenstate 
$|\mb{p} \rangle$. 
Replacing 
the field $\chi_i$ and the relativistic energy $\omega_{\mb{p}_i}$ with a particle position
$\mb{x}_i$ and the non-relativistic one 
$E_{\mb{p}_i}=\mb{p}^2_i/2m_i$, we have the plane wave $\psi_{\mb{p}_i} (\mb{x}_i,t)=e^{-iE_{\mb{p}_i} t} \langle \mb{x}_i|\mb{p}_i \rangle = e^{i(\mb{p}_i\cdot \mb{x}_i-E_{\mb{p}_i} t)}/(2\pi)^{3/2}$. Further adopting the density matrix propagator in Eq.~\eqref{K-NR}, we rederive the non-relativistic scattering supermatrix given in Eq.~\eqref{eq:slashedM-NR}. 

The density matrix propagator $\slashed{K}$ in the field configuration basis is
\be
\slashed{K}(\chi_{1,f},\chi_{2,f},\ul{\chi}_{1,f},\ul{\chi}_{2,f},t_f|\chi_{1,0},\chi_{2,0},\ul{\chi}_{1,0},\ul{\chi}_{2,0},t_0) =\int D\chi_1 D\chi_2 D\ul{\chi}_1 D\ul{\chi}_2 e^{i (S_0[\boldsymbol{\chi}] - S_0[\ul{\boldsymbol{\chi}}])} e^{i S_{IF}[\boldsymbol{\chi},\ul{\boldsymbol{\chi}}]}
\ee
with the boundary condition of the path integral given by the argument of $\slashed{K}$. 
As in Sec.~\ref{sbsc:NR}, we can perturbatively evaluate the propagator $\slashed{K}$ by expanding $e^{iS_{IF}}$ as a series in the coupling strength $\lambda$. 
The ingredients $M_{F_1}$ and $M_{F_2}$ needed for $\slashed{M}$ up to the second order of $\lambda$ are given by 
\begin{align}
\begin{split}
M_{F_1}(p_1p_2 \to p_1'p_2')
&=-\lambda^2  \int d^4y_1  d^4y_2 F_1(y_1-y_2) 
\\
& 
\times \int d\chi_1 d\chi_2 \, \chi^2_1 (y_1) \chi^2_2 (y_2)  
\Psi^*_{p'_1}(\chi_1,y^0_1) 
\Psi^*_{p'_2}(\chi_2,y^0_2)
\Psi_{p_1}(\chi_1,y^0_1) 
\Psi_{p_2}(\chi_2,y^0_2),
\end{split}
\end{align}
and 
\begin{align}
\begin{split}
M_{F_2}(p_i\ul{p}_j \to p_i'\ul{p}_j')
&=\lambda^2  \int d^4y_1  d^4y_2 F_2(y_1-y_2) 
\\
& 
\times \int d\chi_i d\ul{\chi}_j \, \chi^2_i (y_1) \ul{\chi}^2_j (y_2)  
\Psi^*_{p'_i}(\chi_i,y^0_1) 
\Psi_{\ul{p}'_j}(\ul{\chi}_j,y^0_2)
\Psi_{\ul{p}_i}(\chi_i,y^0_1) 
\Psi^*_{\ul{p}_j}(\ul{\chi}_j,y^0_2),
\end{split}
\end{align}
where the integrals of field variables $\chi_1$ and $\chi_2$ are performed on time slices $y^0_1$ and $y^0_2$, respectively.
The field integrals are evaluated as
\be
\int d\chi \, \chi^2(y) \Psi^*_{p'} (\chi,y^0) \Psi_p (\chi,y^0)=\langle \mb{p}'| \hat{\chi}^2 (y) |\mb{p} \rangle=\frac{e^{-i(p'-p)\cdot y}}{(2\pi)^3\sqrt{\omega_{\mb{p}'}\omega_\mb{p}}},
\ee
where $p\neq p'$ and $\hat{\chi}(y)$ is the solution of free field.
After the field integrals and the remaining integrals of $y_1$ and $y_2$, we have the relativistic version of $M_{F_1}$ and $M_{F_2}$,
\begin{align}
\begin{split}
\label{MF1,2-R}
M_{F_1}(p_1p_2 \to p_1'p_2')  &=- \frac{\lambda^2\delta^4(p_1 + p_2 - p_1' - p_2') }{(2\pi)^2 \sqrt{\omega_{\mb{p}_1} \omega_{\mb{p}_2}\omega_{\mb{p}'_1} \omega_{\mb{p}'_2}}} \left[ \frac{i}{k^2 + m_{\phi}^2} + \frac{D_2}{(k^2 + m_{\phi}^2)^2} +  D_0 \right]_{k =p_1' - p_1},
\\
M_{F_2}(p_i\ul{p}_j \to p_i'\ul{p}_j') & = \frac{\lambda^2\delta^4(p_i' - p_i - (\ul{p}_j'  - \ul{p}_j))}{(2\pi)^2 \sqrt{\omega_{\mb{p}_i}\omega_{\ul{\mb{p}}_j}\omega_{\mb{p}'_i}\omega_{\ul{\mb{p}}'_j}}}\left[\frac{D_2}{(k^2 + m_{\phi}^2)^2} +  D_0 \right]_{k = p_i' - p_i}.
\end{split}
\end{align}
Here, we took the limit $\epsilon \rightarrow 0$ since there is no pole in the denominator because of 
$(p'_i-p_i)^2 + m^2_\phi > 0$. In the non-relativistic limit, the above $M_{F_1}$ and $M_{F_2}$ are consistent with our non-relativistic results, Eqs.~\eqref{MF1-final} and \eqref{MF2}. 

At this stage, we can discuss the Lorentz invariance of the resulting scattering supermatrix elements. In unitary $S$-matrix theory, scattering is described via scattering amplitudes, which are covariant under Lorentz transformations, and appropriate combination of the matrix elements with overall energy factors are Lorentz invariant functions of the external momenta. We remind the reader of the standard results and notation in Appendix~\ref{app:lorentz}. In the CQ results shown here, it is easy to check that $p_i \rightarrow \Lambda p_i$ and similarly the underlined (bra) and primed (final state) variables, we have that
\begin{align}
\sqrt{\omega_{\mb{p}_1} \omega_{\mb{p}_2}\omega_{\mb{p}'_1} \omega_{\mb{p}'_2}}M_{F_1},  \ \ \ \sqrt{\omega_{\mb{p}_i} \omega_{\ul{\mb{p}}_j}\omega_{\mb{p}'_i} \omega_{\ul{\mb{p}}'_j}}M_{F_2}
\end{align}
are both invariant. These are just the straightforward analogues of the $S$-matrix element transformations. In particular, this means that the full supermatrix element is Lorentz invariant. In this sense, the $\chi_1 \chi_2 \rightarrow \chi_1 \chi_2$ scattering in the CQ model is consistent with Lorentz symmetry.

Finally, let us look at the transition probability in the full relativistic setting, 
\begin{align}
\begin{split}
\label{eq:P-R}
& P(p_1 p_2 \to p_1' p_2') = \left| M_{F_1}(p_1p_2 \to p_1'p_2') \right|^2   + \left[ M_{F_2}(p_1p_2 \to p_1'p_2') \right]^2 \\
& = \frac{VT}{(2\pi)^4} \frac{\lambda^4\delta^4(p_1' + p_2' - p_1 - p_2) }{(2\pi)^4 \prod_{i=1,2} \omega_{\mb{p}_i}\omega_{\mb{p}'_i}} \left[ \left( \frac{1}{(p_1' - p_1)^2 + m_{\phi}^2} \right)^2 + \left( \frac{D_2}{[(p_1'- p_1)^2 + m_{\phi}^2]^2} +  D_0 \right)^2 \right] \\
& + \frac{VT}{(2\pi)^4} \frac{\lambda^4 \delta^4(p_1' - p_1 - (p_2' - p_2)) }{(2\pi)^4 \prod_{i=1,2} \omega_{\mb{p}_i}\omega_{\mb{p}'_i}} \left(   \frac{D_2}{[(p_1' - p_1)^2 + m_{\phi}^2]^2} +  D_0 \right)^2 
\end{split}
\end{align}
where $\delta^4(0)=VT/(2\pi)^4$. 
The first term is proportional to $\delta^4(p_1' + p_2' - p_1 - p_2)$ and hence this follows the four-momentum conservation, on the other hand, the last term does not. As discussed above, the $\chi_1 \chi_2 \rightarrow \chi_1 \chi_2$ scattering process in the CQ model is consistent with the Lorentz symmetry, but it does not follow the four-momentum conservation law.

\section{Comparison to astrophysical scattering}
\label{sec:example}

\begin{figure}[b]
\begin{tikzpicture}
\draw [fill,color=gray] (-2,0) circle (1);
\node at (-2,-1.5) {$M$};
\draw plot [smooth,dotted] coordinates {(1,-3) (1,-1) (.5,1) (-.5,3)};
\draw [fill,color=gray] (.5,1) circle (.1);
\node at (1,1) {$m$};
\draw [->,thick] (1,-3) -- (1,-2.5); 
\node at (1.5,-2.7) {$\mb{p}$};
\draw [->,thick] (-.25,2.5) -- (-.5,3); 
\node at (0,3) {$\mb{p}'$};
\end{tikzpicture}
\caption{Heavy-light scattering, for example a spacecraft of mass $m$ performing a slingshot maneuver around a planet of mass $M$.}
\label{fig:slingshot}
\end{figure}
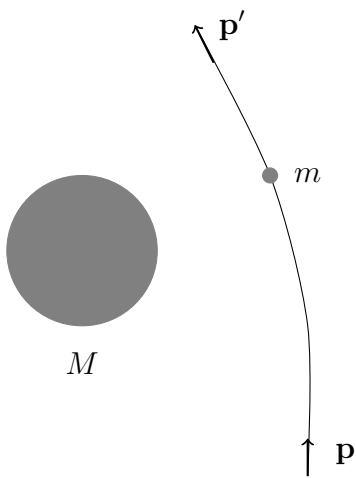

To compare this model to the real world, let's consider scattering a small body, like a spacecraft of mass $m$, against a large body, like a planet of mass $M$, as in a slingshot maneuver to accelerate a spacecraft. See Fig. \ref{fig:slingshot}. We choose this because it is a clear scattering problem, and because there is extensive and precise data from real spacecraft. Another option could be to analyze asteroid scattering, for example.

We can analyze this problem in the non-relativistic setting for simplicity. To match to the Newton potential, we identify the Yukawa coupling as
\be
\lambda = \sqrt{G_N} m M.
\ee
Also, we send the Yukawa mass $m_{\phi} \to 0$. Since $M \gg m$, the reduced mass $\mu \approx m$. We will work in the rest frame of $M$. Let $\mb{p} = p \hat{\mb{z}}$ be the initial momentum of the small object. We now compute the effects of the three terms in Eq. \eqref{dsigma-nr}. 

The first term,
\begin{align}
\frac{d\sigma}{d^3\mb{p}' d^3\mb{P}'} \supset \frac{(2\pi)^2}{u} \frac{\lambda^4}{m^2 M^2} \delta^4(p' + P' - p - P)  \left[ \frac{1}{(\mb{p}' - \mb{p})^4} + \left( \frac{D_2}{(\mb{p}' - \mb{p})^4} +  D_0 \right)^2 \right],
\end{align}
has the usual overall energy-momentum conservation delta function and thus can be analyzed like a standard scattering problem. First we integrate $\delta^3(\mb{p}' + \mb{P}' - \mb{p}) d^3\mb{p}' d^3\mb{P}' = d^3\mb{p}'$, and set $\mb{P}' := \mb{p} - \mb{p}'$ everywhere. We then integrate the energy delta function $\delta(E_{\mb{p}'} + E_{\mb{P}'} - E_{\mb{p}}) d^3\mb{p}' \approx m p d\Omega$, using $m/M \ll 1$, and set $|\mb{p}'| = |\mb{p}| = p$ everywhere. In particular, we then have $\mb{p}' = p \hat{\mb{n}}$ where $\hat{\mb{n}}$ is a unit three-vector. The cross-section becomes
\begin{align}
\label{dsigma-1}
d\sigma \supset 4\pi^2 G_N^2 M^2 m^4  \left[  \frac{1}{16 p^4 \sin^4(\theta/2)} + \left( \frac{D_2}{16 p^4 \sin^4(\theta/2)} +  D_0 \right)^2 \right]d\Omega,
\end{align}
where $\hat{\mb{z}} \cdot \hat{\mb{n}} = \cos \theta$ and we used $u = p/\mu$. The first term here is just the ordinary formula for Rutherford/Newtonian scattering. The subsequent terms involving $D_{0,2}$ are new contributions from the CQ theory. We comment on their effects below.

Now consider the second term in Eq. \eqref{dsigma-nr}, 
\begin{align}
\frac{d\sigma}{d^3\mb{p}' d^3\mb{P}'} \supset \frac{(2\pi)^2}{u} \frac{\lambda^4}{m^2 M^2} \delta^4(p' - p - (P' - P)) \left(  \frac{D_2}{(\mb{p}' - \mb{p})^4} +  D_0 \right)^2.
\end{align}
The phenomenology of this term is very different. The energy-momentum delta function does not enforce the usual conservation rule but rather says that both particles receive \emph{equal} kicks, as opposed to equal and opposite kicks. These kicks come from the stochastic field. The three-momentum delta function now correspondingly says that $\mb{P}' = \mb{p}' - \mb{p}$, and the corresponding integral gives $\delta^3(\mb{p}' - \mb{p} - \mb{P}') d^3\mb{p}' d^3\mb{P}' = d^3\mb{p}'$ as before. The energy delta now enforces
\be
0 = E_{\mb{p}'} - E_{\mb{p}} - E_{\mb{P}'} \approx \frac{|\mb{p}'|^2 - |\mb{p}|^2}{2m},
\ee
where we used $m \ll M$, which here means we neglect the energy change in the large mass. So this again sets $\mb{p}' = p \hat{\mb{n}}$ where $\hat{\mb{n}}$ is any unit three-vector. We can then reduce the remaining volume element in the same way as the previous paragraph, and we obtain
\begin{align}
\label{dsigma-2}
d\sigma \supset  4\pi^2 G_N^2 M^2 m^4 \left(   \frac{D_2}{16 p^4 \sin^4(\theta/2)} +  D_0 \right)^2 d\Omega.
\end{align}
This is identical to the CQ contribution in Eq. \eqref{dsigma-1}. We emphasize that this is only the case in our approximation that $m \ll M$.

Putting these results together, we obtain the differential cross-section
\begin{align}
\frac{d\sigma}{d\Omega} = 4\pi^2 G_N^2 M^2 m^4 \left[ \frac{1}{16 p^4 \sin^4(\theta/2)} + 2\left( \frac{D_2}{16 p^4 \sin^4(\theta/2)} +  D_0 \right)^2 \right].
\end{align}
Again, the first term here is the usual formula obtained for Newtonian/Rutherford scattering. Thus this equation represents a deviation from the usual predictions of classical gravitational scattering, encoded in the second term. This term comes from the noise added in the CQ model. In particular, its total contribution is lower-bounded by the CQ diffusion-decoherence constraint $D_0 D_2 \geq 1$. Moreover, since this term has a nearly identical angular dependence as the usual Newtonian term, this appears to change gravitational scattering distributions at $\Oi(1)$, which is in conflict with basic observations of these kinds of processes. We emphasize again that this conclusion is obtained only in the simplest case of $D_{0,2}$ constant, although it holds for any values of these constants subject to the CQ constraint $D_0 D_2 \geq 1$.

\section{Comments and outlook}
\label{sec-conclusions}

It is of elementary interest to know if and how relativistic quantum fields can couple to classical ones. In the context of gravity, while our experience with the other gauge fields of nature strongly suggests that the metric should be similarly quantized, it is also of elementary interest to determine if there is any self-consistent way that the metric can be treated as a classical stochastic variable. 

In this paper, we have analyzed the recent classical-quantum (``CQ'') proposal of Oppenheim \emph{et al.}, which provides an attempt at an affirmative, constructive answer to both of these questions. Perhaps surprisingly, we have found that a version of tree-level scattering theory can be formulated in an appropriate limit of this model. As advertised by Oppenheim \emph{et al.}, the scattering probabilities are Lorentz invariant. Counter-intuitively, the probabilities do not obey energy-momentum conservation. These two facts are consistent because the model is fundamentally an open system, and there is no analogue of Noether's theorem. Moreover, although the model is an open system and thus not unitary, total probability is still conserved, so the results appear to be self-consistent at this basic level.

Quantitatively, we have analyzed scattering of two massive objects in this CQ framework, both relativistic and non-relativistic. Taken at face value, the non-relativistic result is in direct conflict with scattering of astrophysical objects. The scattering probabilities over-predict small-angle scattering, as well as the total cross-section, by an $\Oi(1)$ amount. This may suggest that the detailed construction is ruled out. 

There are three main caveats to this conclusion. The obvious one is that we are working with a simple Yukawa interaction, not full general relativity, although for our non-relativistc calculations this should not pose any problems. Another is that we are only working in the simplest case where the diffusion and decoherence functions $D_{0,2}$ are constants. We make no claims about what happens when these are allowed to be field-dependent functionals. Perhaps the trickiest caveat is that we are assuming that we can use this formalism treating a large composite object as a point mass. In ordinary classical or perturbative quantum gravity, this works automatically because of the Gauss law. Here, we should really do a more careful study of the renormalization of the model to understand how to go from microscopic interactions to macroscopic interactions. This is an interesting direction that we leave to future work.

Generally speaking, our conclusion is that the Oppenheim \emph{et al.} construction is an interesting and non-trivial attempt to couple relativistic quantum and classical systems. At a structural level it seems to produce basic features that any such coupling should have, namely Lorentz invariance and conservation of probability. However, as a model of quantum gravity, the simplest case ($D_{0,2} = \rm{constant}$) seems to be in conflict with the real world. Moreover, the lack of friction in the diffusive evolution of the classical fields may be fundamentally problematic. These issues warrant further study.

\section*{Acknowledgements}

We thank Isaac Layton, Jonathan Oppenheim, Geoff Penington, Shivaji Sondhi, Jacob Taylor, and Zach Weller-Davies for valuable discussions. D.C. is supported by the U.S. DOE, Office of High Energy Physics, under Contract No. DEAC02-05CH11231, by DOE Quantum Information Science Enabled Discovery (QuantISED) for High Energy Physics grant KA2401032, and by the Heising-Simons Foundation grant 2023-4467 ``Testing the Quantum Coherence of Gravity''. A.M. is supported by JSPS KAKENHI (Grant No.~JP23K13103 and No.~JP23H01175).

\newpage

\bibliographystyle{utphys-dan.bst}
\bibliography{cq-grav}

\newpage

\appendix

\section{Classical equations of motion}
\label{app:C}

In this appendix, we consider the time evolution of classical field $\phi$, averaging over the quantum matter. In particular, we want to show in what sense the CQ framework produces a ``semiclassical'' equation of motion where $\phi$ is sourced by expectation values of the quantum matter, e.g., by $\lambda \braket{\boldsymbol{\chi}^2}$. To this end, we begin with the following CQ path integral,
\be
\varrho(\boldsymbol{\chi}_f,\ul{\boldsymbol{\chi}}_f,\phi_f,\pi_f,t_f) = \int D\boldsymbol{\chi}  D\ul{\boldsymbol{\chi}} D\phi D\pi e^{I_{CQ}[\boldsymbol{\chi},\ul{\boldsymbol{\chi}},\phi,\pi]} \varrho( \boldsymbol{\chi}_0,\ul{\boldsymbol{\chi}}_0,\phi_0,\pi_0,t_0),
\label{eq:rho-app}
\ee
where $D\boldsymbol{\chi}=D\chi_1 D\chi_2$, 
$D\ul{\boldsymbol{\chi}}=D\ul{\chi}_1 D\ul{\chi}_2$, and the CQ action is 
\begin{align}
\begin{split}
& I_{CQ}[\boldsymbol{\chi}, \underline{\boldsymbol{\chi}}, \phi,\pi]  =  i S_0 [\boldsymbol{\chi}]- i S_0 [\ul{\boldsymbol{\chi}}] 
+i \lambda \int d^4x  (\boldsymbol{\chi}^2 - \ul{\boldsymbol{\chi}}^2 ) \phi +\ln (\delta[\dot{\phi}-\pi])
\\ 
& -\frac{D_0 \lambda^2}{2} \int d^4x  (\boldsymbol{\chi}^2 - \ul{\boldsymbol{\chi}}^2)^2
-\frac{1}{2D_2} \int d^4x \Big[\dot{\pi}-(\nabla^2-m^2_\phi) \phi -\frac{\lambda}{2} (\boldsymbol{\chi}^2 +\ul{\boldsymbol{\chi}}^2) \Big]^2. 
\label{eq:Iapp}
\end{split}
\end{align}
This CQ action is computed from Eq.~\eqref{eq:I} by adopting the Yukawa interaction $S_\text{int}$ of Eq.~\eqref{eq:Yukawa}. 

Since the the term with $D_2$ generates the diffusive and stochastic evolution of the classical field $\phi$, we expect that its time evolution is governed by Langevin equations. To see this explicitly, we can use the identity
\be
1 = \int D\xi \delta\Big[\dot{\pi}-(\nabla^2-m^2_\phi) \phi-\frac{\lambda}{2} (\boldsymbol{\chi}^2 +\ul{\boldsymbol{\chi}}^2)- \sqrt{\frac{D_2}{2}} \xi\Big].
\ee
Here $\xi = \xi(x)$ is again a white noise random variable [c.f. Eq. \eqref{eq:langevin-example}], which generates the diffusive evolution of the classical field. Concretely, we can use this identity and Eq. \eqref{eq:Iapp} to rewrite Eq. \eqref{eq:rho-app} as follows:
\begin{align}
\begin{split}
\label{eq:app-full-integral}
&\varrho(\boldsymbol{\chi}_f,\ul{\boldsymbol{\chi}}_f,\phi_f,\pi_f,t_f) 
\\
& \quad 
= \int D\xi e^{-\int d^4x \, \xi^2} \int D\boldsymbol{\chi}  D\ul{\boldsymbol{\chi}} D\phi D\pi \delta[\dot{\phi}-\pi]\delta \Big[\dot{\pi}-(\nabla^2-m^2_\phi) \phi -\frac{\lambda}{2} (\boldsymbol{\chi}^2 +\ul{\boldsymbol{\chi}}^2)-\sqrt{\frac{D_2}{2}} \xi\Big]\, 
\\
& \quad 
\times e^{ i S_0 [\boldsymbol{\chi}]- i S_0 [\ul{\boldsymbol{\chi}}] 
+i \lambda \int d^4x  (\boldsymbol{\chi}^2 - \ul{\boldsymbol{\chi}}^2 ) \phi }
e^{-\frac{D_0 \lambda^2}{2} \int d^4x  (\boldsymbol{\chi}^2 - \ul{\boldsymbol{\chi}}^2)^2}
\varrho( \boldsymbol{\chi}_0,\ul{\boldsymbol{\chi}}_0,\phi_0,\pi_0,t_0).
\end{split}
\end{align}
The delta functionals condition the classical trajectory of $(\phi,\pi)$ on 
\be
\label{eq:Lgvn1}
\dot{\phi}=\pi, \quad 
\dot{\pi}=(\nabla^2-m_\phi ^2) \phi+\frac{\lambda}{2} (\boldsymbol{\chi}^2 +\ul{\boldsymbol{\chi}}^2)+\sqrt{\frac{D_2}{2}} \xi,
\ee 
and these are the Langevin equations of classical field. 

With this rewriting of the path integral, it is easy to see how the classical field evolves. Consider the expectation value of $\dot{\pi}$, for example. This is given by
\begin{align}
\begin{split}
\label{app:dotpi}
\braket{ \dot{\pi} } & =  \int D\boldsymbol{\chi}  D\ul{\boldsymbol{\chi}} D\phi D\pi\, e^{I_{CQ}[\boldsymbol{\chi},\ul{\boldsymbol{\chi}},\phi,\pi]} \times \dot{\pi} \times \varrho( \boldsymbol{\chi}_0,\ul{\boldsymbol{\chi}}_0,\phi_0,\pi_0,t_0)  \\
&= \int D\xi e^{- \int  \, d^4x \xi^2} \int D\boldsymbol{\chi}  D\ul{\boldsymbol{\chi}} D\phi D\pi\, \delta[\dot{\phi}-\pi] \delta \Big[\dot{\pi}-(\nabla^2-m^2_\phi) \phi -\frac{\lambda}{2} (\boldsymbol{\chi}^2 +\ul{\boldsymbol{\chi}}^2)-\sqrt{\frac{D_2}{2}} \xi\Big] 
\\
&
\times 
e^{ i S_0 [\boldsymbol{\chi}]- i S_0 [\ul{\boldsymbol{\chi}}] 
+i \lambda \int d^4x  (\boldsymbol{\chi}^2 - \ul{\boldsymbol{\chi}}^2 ) \phi }
e^{-\frac{D_0 \lambda^2}{2} \int d^4x  (\boldsymbol{\chi}^2 - \ul{\boldsymbol{\chi}}^2)^2}
\\
& \times \Big\{ (\nabla^2-m^2_\phi) \phi +\frac{\lambda}{2} (\boldsymbol{\chi}^2 +\ul{\boldsymbol{\chi}}^2)+ \sqrt{\frac{D_2}{2}} \xi \Big \} \times \varrho( \boldsymbol{\chi}_0,\ul{\boldsymbol{\chi}}_0,\phi_0,\pi_0,t_0)
\\
&= (\nabla^2-m^2_\phi) \langle \phi \rangle +\lambda \langle \boldsymbol{\chi}^2 \rangle +\sqrt{\frac{D_2}{2}} \langle \xi \rangle.
\end{split}
\end{align}
The first line is the usual expression for an expectation value in a path integral. The second line was obtained by inserting Eq. \eqref{eq:app-full-integral} and using the delta function to replace $\dot{\pi}$ with the equation of motion. To obtain the third line, we used $\langle \ul{\boldsymbol{\chi}}^2 \rangle=\langle \boldsymbol{\chi}^2 \rangle^*=\langle \boldsymbol{\chi}^2 \rangle$. Note that we should take the  boundary condition $\boldsymbol{\chi}=\ul{\boldsymbol{\chi}}$ in the upper limit of the $D\boldsymbol{\chi} D\ul{\boldsymbol{\chi}}$ path integral, which is called the closed-time-path condition.

The result in Eq.~\eqref{app:dotpi} tells us that the classical field is excited by the quantum field through the source term $\lambda \langle \boldsymbol{\chi}^2 \rangle$, i.e., by the expectation value of the quantum matter. It is also excited by the white noise $\xi$, which generally in the main text we have taken to have zero mean $\braket{\xi} = 0$.

\section{Feynman-Vernon derivations}
\label{app:fv}

Here we derive the Feynman-Vernon path integrals used in the main text. 
Our starting point is the path integral for giving the influence functional,
\be
\label{app:SIF}
e^{iS_0 [\boldsymbol{\chi}]-iS_0[\ul{\boldsymbol{\chi}}] + iS_{IF}[\boldsymbol{\chi},\ul{\boldsymbol{\chi}}]}= \int D\phi D\pi e^{I_{CQ}[\boldsymbol{\chi},\ul{\boldsymbol{\chi}}, \phi, \pi]} \mathcal{P}[\phi_0,\pi_0,t_0],
\ee
where $S_0 [\boldsymbol{\chi}]$ the free action of massive scalar fields $\boldsymbol{\chi}=(\chi_1, \chi_2)$ and $\mathcal{P}[\phi_0,\pi_0,t_0]$ is the initial distribution of classical field. The CQ action is 
\begin{align}
\begin{split}
I_{CQ}[\mb{x}_1,\mb{x}_2, \ul{\mb{x}}_1,\ul{\mb{x}}_2, \phi, \pi]
& 
= \ln (\delta[\pi-\dot{\phi}])+ i S_0 [\boldsymbol{\chi}]- i S_0 [\ul{\boldsymbol{\chi}}] 
+i \int d^4x  (J - \ul{J} ) \phi 
\\
&
-\frac{D_0 }{2} \int d^4x  (J - \ul{J})^2
-\frac{1}{2D_2} \int d^4x \Big[\dot{\pi}-(\nabla^2-m^2_\phi) \phi -\frac{1}{2} (J + \ul{J} ) \Big]^2
\end{split}
\end{align}
with the currents of quantum particles,
\begin{align}
\begin{split}
\label{app:J}
J (x)= \sum_i J_i (x), \quad J_i(x) = \lambda \chi^2_i (x), \quad 
\ul{J} (x)= \sum_i \ul{J}_i (x), \quad \ul{J}_i (x) =\lambda \ul{\chi}^2_i (x).
\end{split}
\end{align}
Substituting this CQ action into Eq.~\eqref{app:SIF}, we have the following expression of the influence functional 
$S_\text{IF}$, 
\begin{align}
\begin{split}
e^{iS_\text{IF}}
&= \int D\phi D\pi \delta[\pi-\dot{\phi}] e^{i\int d^4x  (J - \ul{J})\phi
-\frac{D_0 }{2} \int d^4x  (J - \ul{J})^2-\frac{1}{2D_2} \int d^4x \big[\dot{\pi}-(\nabla^2-m^2_\phi) \phi -\frac{1}{2} (J  + \ul{J}) \big]^2} \mathcal{P}[\phi_0,\pi_0,t_0].
\end{split}
\end{align}
Using the identity, 
\be
\int D\xi \delta\Big[\dot{\pi}-(\nabla^2-m^2_\phi) \phi-\frac{1}{2} (J + \ul{J})-\xi\Big]=1,
\ee
we then arrive at 
\begin{align}
\begin{split}
e^{iS_\text{IF}}
&= \int D\xi e^{-\frac{1}{2D_2} \int d^4x \xi^2 -\frac{D_0 }{2} \int d^4x  (J- \ul{J})^2}
\\
&\times 
\int D\phi D\pi \delta[\pi-\dot{\phi}] \delta \big[\dot{\pi}-(\nabla^2-m^2_\phi) \phi-\frac{1}{2} (J + \ul{J})-\xi\big] 
e^{i \int d^4x (J  - \ul{J} ) \phi} \mathcal{P}[\phi_0,\pi_0,t_0].
\end{split}
\end{align}
Let us evaluate the path integral with respect to $\phi$ and $\pi$. 
The delta functionals $\delta[\pi-\dot{\phi}]$ and $ \delta\left[\dot{\pi}-(\nabla^2-m^2_\phi) \phi-\frac{1}{2}(J+ \ul{J} )-\xi\right] $ suggest that the classical field follows the Langevin equations, 
\be
\dot{\phi}=\pi, \quad 
\dot{\pi}=(\nabla^2-m_\phi ^2) \phi+\frac{1}{2} (J + \ul{J})+\xi,
\ee
where the white noise $\xi$ obeys the Gaussian statistics, 
\be
\int D \xi e^{-\frac{1}{2D_2} \int d^4x 
\, \xi^2} \xi(x)=0, \quad \int D \xi e^{-\frac{1}{2D_2} \int d^4x 
\, 
\xi^2} \xi(x) \xi(y)= D_2 \delta^4(x-y). 
\ee
The initial condition of the Langevin equations is determined by the initial distribution $\mathcal{P}[\phi_0,\pi_0,t_0]$. Here, we adopt the initial vacuum state of classical field with $\mathcal{P}[\phi_0,\pi_0,t_0]=\delta[\phi(t_0)] \delta[\pi(t_0)]$ at the initial time $t_0$. This means that the classical field $\phi$ is initially zero and it is sourced by the currents $J_i, \ul{J}_i$ of quantum particles and the white noise $\xi$. Explicitly, the solution $\phi$ is 
\begin{align}
\label{app:sol}
\phi (x)=\int d^4y \, G_R (x-y) \left[ \frac{1}{2} \{ J(y) + \ul{J} (y)\}+\xi (y) \right],
\end{align}
where
\begin{align}
G_R (x-y)=\int \frac{d^4k}{(2\pi)^4} \frac{e^{ik\cdot(x-y)}}{-(k^0+i\epsilon)^2+\mb{k}^2+m_{\phi}^2},
\end{align}
is the retarded Green's function. It satisfies $(-\partial^2_x +m_{\phi}^2) G_R(x-y)=\delta^4(x-y)$ and  $G_R(x-y)=0$ for $x^0-y^0<0$. Hence, after performing the integration with respect to $\phi$ and $\pi$, we just should replace  $\phi$ with the solution given in Eq.~\eqref{app:sol}. Following this rule, we can get the exact form of $e^{iS_{IF}}$ as
\begin{align}
\begin{split}
e^{iS_{IF}}
&= \int D\xi e^{-\frac{1}{2D_2} \int d^4x \xi^2 -\frac{D_0 }{2} \int d^4x  (J - \ul{J})^2}
\\
&\times 
\int D\phi D\pi \delta[\pi-\dot{\phi}] \delta \big[\dot{\pi}-(\nabla^2-m^2_\phi) \phi-\frac{1}{2} (J + \ul{J})-\xi\big] 
e^{i\int d^4x  (J - \ul{J}) \phi} \mathcal{P}[\phi_0,\pi_0,t_0]
\\
&= \int D\xi e^{-\frac{1}{2D_2} \int d^4x \xi^2 -\frac{D_0}{2} \int d^4x  (J - \ul{J})^2+i \int d^4x d^4y (J- \ul{J} )G_R  \left[\frac{1}{2}(J+ \ul{J} ) + \xi \right ] } 
\\
&\times 
\int D\phi D\pi \delta[\pi-\dot{\phi}] \delta \big[\dot{\pi}-(\nabla^2-m^2_\phi) \phi-\frac{1}{2} (J + \ul{J})-\xi\big] \delta[\phi(t_0)]\delta[\pi(t_0)]
\\
&= e^{-\frac{D_0 }{2} \int d^4x  (J - \ul{J})^2+i\frac{1}{2} \int d^4x d^4y (J- \ul{J})G_R(J+ \ul{J}) } \int D\xi e^{-\frac{1}{2D_2} \int d^4x \xi^2 +i \int d^4x d^4y (J- \ul{J} ) G_R \, \xi } 
\\
&= e^{-\frac{D_0 }{2} \int d^4x  (J - \ul{J})^2+i\frac{1}{2} \int d^4x d^4y (J- \ul{J})G_R(J+ \ul{J}) -\frac{D_2 }{2}\int d^4x d^4y (J- \ul{J} ) G_C (J- \ul{J} )} 
\end{split}
\end{align}
In the second equality, the field 
$\phi$ of the exponential was replaced with the solution given in Eq.~\eqref{app:sol}, and in the third equality the 
$\phi$ and 
$\pi$ path integrals were performed. In the fourth equality, we did the Gaussian integration with respect to $\xi$. The two point function $G_C =G_C (x-y)$ is 
\begin{align}
G_{C} (x-y)
&= \int d^4 z \,  G_{R}(x-z)G_{R}(y-z)
\nonumber 
\\
&=\int \frac{d^4 k}{(2\pi)^4} \, \frac{e^{ik\cdot(x-y)}}{[-(k^0+i\epsilon)^2+\mb{k}^2+m_\phi^2][-(k^0-i\epsilon)^2+\mb{k}^2+m_\phi^2]}.
\label{app:GC2}
\end{align}
The influence functional $S_{IF}$ is thus 
\be
\label{app:IF_full}
iS_{IF} = -\frac{1}{2} \int d^4 x d^4y J(x) F_1 (x-y) J (y)-2J(x) F_2 (x-y) \ul{J} (y)+\ul{J}(x) F^*_1 (x-y) \ul{J} (y),
\ee
where 
\begin{align}
\begin{split}
&F_1 (x-y)=D_2 G_{C} (x-y)-\frac{i}{2} [G_{R} (x- y)+G_{R} (y-x)]+D_0 \delta^4(x-y), 
\\
&F_2 (x-y)= D_2 G_{C} (x-y)
+\frac{i}{2} [G_{R} (x-y)-G_{R} (y-x)]+D_0 \delta^4(x-y).
\end{split}
\end{align}
Since the current is $J=\sum_i J_i$, in the influence functional, we have the self interaction term  $J^2_i$ and $\ul{J}^2_i$. For the scattering process of interest, we pick up the cross terms $J_i J_j$, $\ul{J}_i \, \ul{J}_j$ and $J_i \, \ul{J}_j $ with $i\neq j$. The influence functional only with the cross terms is 
\be
\label{app:IF}
iS_{IF}
= 
- \int d^4 x d^4y J_1 (x) F_1 (x-y) J_2 (y)+\ul{J}_1 (x) F^*_1 (x-y) \ul{J}_2 (y)-\sum_{i\neq j} J_i(x) F_2 (x-y) \ul{J}_j (y). 
\ee
Substituting the form of each current \eqref{app:J} for this, we can produce the influence functional given in Eq.~\eqref{IF-r}. 

To get the non-relativistic limit, note that the above procedure for deriving Eq.~\eqref{app:IF} does not rely on the concrete form of the currents. To get the non-relativistic version \eqref{IF-nr}, we can substitute the current $J_i(x)=\frac{\lambda}{m_i} \delta^3 (\mb{x}-\mb{x}_i)$ for the influence functional. The non-relativistic propagators are reduced by noting that the momentum transfer $k^{\mu} = p'^{\mu} - p^{\mu}$, so $k^0 \approx (\mb{p}'^2-\mb{p}^2)/2m \ll |\mb{k}|$. This means we can drop the $k^0$ terms in the denominators, which in particular produces $\int dk^0 e^{i k^0 (x^0 - y^0)} = 2\pi \delta(x^0 - y^0)$, which is why the non-local $S_{IF}$ collapses to a time-local expression.

As discussed here and in the main text, we are only focusing on terms of the form $J_1 J_2$, rather than allowing for any self-interactions like $J_1 J_1$. This is because we are looking at tree-level scattering to lowest order in perturbation theory. To this order, the self-interaction terms only contribute as loop diagrams in the forward scattering process, so as long as we consider non-forward scattering this is sufficient. However, these self-energy diagrams are definitely of importance in the CQ model, and will behave differently than in a unitary QFT. In particular, the self-energy of the matter will necessarily have finite width coming from the $D_{0,2}$ interactions. Understanding how to deal with these loops via some kind of renormalization procedure is a very interesting problem which we leave to future work (see, e.g.,~\cite{Grudka:2024llq} for some further comments).


\section{Lorentz invariance of scattering supermatrix}
\label{app:lorentz}

In Sec.~\ref{subsec:rscattering}, we have observed the Lorentz invariance of the scattering supermatrix elements in our CQ model. Here, to help our understanding of why this invariance appears, we discuss the Lorentz invariance of the scattering supermatrix in the quantum counterpart of our CQ model, where all the fields including $\phi$ are quantized. In particular, one might have wondered by the classical $\phi$ field does not provide a reference frame; we show here why the analogous fully quantum process does not provide a reference frame. This also gives us the opportunity to review the basic transformation properties of transition probabilities, and show how Lorentz-invariant probabilities that do not preserve four-momentum can arise once one drops the constraint of a unitary $S$-matrix.

In the CQ calculations in the main text, we defined a channel $\mathcal{M}$ which gives time evolution on just the $\boldsymbol{\chi}$ fields, see Eq. \eqref{eq:slashM}. This was defined by assuming an initial $\phi$ distribution $\mathcal{P}[\phi,\pi] = \delta[\phi] \delta[\pi]$ and averaging over the final $\phi$ configurations. The analogous channel in the fully quantized setting is
\be
\mathcal{M} \left[ \ket{\alpha} \bra{\ul{\alpha}} \right]
=\text{Tr}_{\phi} 
\Big[S \ket{\alpha} \bra{\ul{\alpha}} \otimes \ket{0} \bra{0}_{\phi} S^\dagger \Big],
\ee
where $S=\lim_{T\rightarrow \infty} e^{iH_0 T} e^{-i2HT}e^{-iH_0 T} $ is the usual S-operator defined with the full Hamiltonian $H$ and the free Hamiltonian $H_0$. The scattering supermatrix element computed from the channel $\mathcal{M}$ is given by the same formula we used in the CQ case, viz.
\be
\slashed{M}_{\alpha \ul{\alpha} \to \beta \ul{\beta}} = \braket{\beta | \mathcal{M} \left[ \ket{\alpha} \bra{\ul{\alpha}} \right] | \ul{\beta}}.
\label{app:element}
\ee
In the following, we will analyze how $\slashed{M}_{\alpha \ul{\alpha} \to \beta \ul{\beta}}$ acquires Lorentz invariance, despite the fact that we are tracing out the final $\phi$ states. In particular this means that the $\phi$ field does not generate some preferred frame for the $\boldsymbol{\chi}$ fields, even though $\phi$ can be excited. The proof does not rely on perturbation theory.

Let $U = V \otimes W$ be a Poincar\'e transformation $U = U(\Lambda,a)$. Following the standard formulation of scattering theory presented in~\cite{weinberg1995quantum}, we assume that this factors into a Poincar\'e transformation $V$ acting on the asymptotic $\boldsymbol{\chi}$ states and a Poincar\'e transformation $W$ acting on the asymptotic $\phi$ states. The vacuum of $\phi$ is invariant
\be
\label{eq:vac_inv}
W \ket{0}_{\phi} = \ket{0}_{\phi},
\ee
and the full transformation commutes with the $S$-matrix
\be
\label{eq:S-com}
[S,U] = 0.
\ee
Using these, we can show that $\mathcal{M}$ is covariant under $V$:
\begin{align}
\begin{split}
\mathcal{M} [V \ket{\alpha} \bra{\ul{\alpha}} V^\dagger ]
&
=\text{Tr}_\phi \Big[S \Big \{ V  \ket{\alpha} \bra{\ul{\alpha}} V^\dagger \otimes \ket{0}\bra{0}_{\phi} \Big \} S^\dagger \Big]
\\
&
=\text{Tr}_\phi \Big[S  \Big \{V  \ket{\alpha} \bra{\ul{\alpha}} V^\dagger\otimes W \ket{0}\bra{0}_{\phi} W^\dagger \Big \} S^\dagger \Big]
\\
&
=\text{Tr}_\phi \Big[U \, S  \ket{\alpha} \bra{\ul{\alpha}} \otimes \ket{0}\bra{0}_{\phi}  S^\dagger \, U^\dagger \Big]
\\
&
=V \text{Tr}_\phi \Big[ W  S  \ket{\alpha} \bra{\ul{\alpha}} \otimes \ket{0}\bra{0}_{\phi}  S^\dagger  W^\dagger \Big]V^\dagger 
\\
&
=V\text{Tr}_\phi \Big[ S   \ket{\alpha} \bra{\ul{\alpha}} \otimes \ket{0}\bra{0}_{\phi}  S^\dagger \Big]V^\dagger
\\
&
=V\mathcal{M} [\ket{\alpha} \bra{\ul{\alpha}} ]V^\dagger.
\label{app:Minv}
\end{split}
\end{align}
The first line is a definition, the second used Eq.~\eqref{eq:vac_inv}, and to get from the fourth to fifth lines we used that $W$ only acts on the $\phi$ Hilbert space so it cycles under the partial trace.

We can be a bit more explicit about the implications of Eq.~\eqref{app:Minv}, which is instructive for comparing to the CQ model. Since we are interested in scalar fields $\boldsymbol{\chi}=(\chi_1, \chi_2)$, we can take the Fock states $|\alpha \rangle$ and $|\beta \rangle$ of $\boldsymbol{\chi}$ particles with definite momenta $\alpha = \mb{p}_1, \mb{p}_2, \ldots$ and $\beta = \mb{p}_1',\mb{p}_2', \ldots$, respectively. In our scattering process, the following matrix element is relevant, 
\be
\slashed{M}_{p_1 p_2 \ul{p}_1 \ul{p}_2 \to p'_1 p'_2 \ul{p}'_1 \ul{p}'_2} = \braket{\mb{p}'_1 \mb{p}'_2 | \mathcal{M} \left[ \ket{\mb{p}_1 \mb{p}_2} \bra{\ul{\mb{p}}_1 \ul{\mb{p}}_2} \right] | \ul{\mb{p}}'_1 \ul{\mb{p}}'_2}.
\label{app:22_element}
\ee
These states transform as usual~\cite{weinberg1995quantum}, 
\be
V(\Lambda,a) |\mb{p}_1 \mb{p}_2 \rangle=\sqrt{\frac{(\Lambda p_1)^0 (\Lambda p_2)^0}{p^0_1 p^0_2}} e^{-i(\Lambda p_1 + \Lambda p_2)^\mu  a_\mu }|\boldsymbol \Lambda \mb{p}_1 \boldsymbol \Lambda \mb{p}_2 \rangle.
\label{app:trns_st}
\ee
Using $V\mathcal{M} [\ket{\alpha} \bra{\ul{\alpha}} ]V^\dagger=\mathcal{M} [V \ket{\alpha} \bra{\ul{\alpha}} V^\dagger ]$, or equivalently $\mathcal{M} [\ket{\alpha} \bra{\ul{\alpha}} ]=V^\dagger \mathcal{M} [V \ket{\alpha} \bra{\ul{\alpha}} V^\dagger ]V$ shown in \eqref{app:Minv}, we have 
\begin{align}
\begin{split}
\slashed{M}_{p_1 p_2 \ul{p}_1\ul{p}_2 \to p'_1 p'_2 \ul{p}'_1 \ul{p}'_2} & = \left[\prod_{i={1,2}}\sqrt{ \frac{(\Lambda p'_i)^0 (\Lambda p_i)^0 (\Lambda \ul{p}'_i)^0 (\Lambda \ul{p}_i)^0}{p'{}^0_i p^0_i \ul{p}'{}^0_i \ul{p}^0_i} } \right] \exp \left[ i\sum_{i=1,2}  ( \Lambda p'_i - \Lambda p_i-\Lambda \ul{p}'_i + \Lambda \ul{p}_i )^\mu  a_\mu \right] \\
& \times
 \slashed{M}_{\Lambda p_1 \Lambda p_2 \Lambda \ul{p}_1 \Lambda \ul{p}_2 \to \Lambda p'_1 \Lambda p'_2 \Lambda \ul{p}'_1 \Lambda \ul{p}'_2}. 
\label{app:slashedM_inv}
 \end{split}
\end{align}
This is the generalization of the usual Lorentz invariance of S-matrix elements, which reads
\be
S_{\mb{p}_1 \mb{p}_2 \to \mb{p}_1' \mb{p}_2'} = \left[\prod_{i={1,2}}\sqrt{ \frac{(\Lambda p'_i)^0 (\Lambda p_i)^0}{p'{}^0_i p^0_i } }\right] \exp \left[ i\sum_{i=1,2}  ( \Lambda p'_i - \Lambda p_i)^\mu  a_\mu \right] S_{\boldsymbol{\Lambda} \mb{p}_1 \boldsymbol{\Lambda} \mb{p}_2 \to \boldsymbol{\Lambda}\mb{p}_1' \boldsymbol{\Lambda}\mb{p}_2'}
\ee
Notice, however, the more general condition can mix the bra and ket variables in the energy-momentum conservation exponential.

To be even more explicit, first consider the case with no spacetime translation $a^\mu=0$. Then Eq.~\eqref{app:slashedM_inv} implies
\begin{align}
\begin{split}
& \Big[\prod_{i}\sqrt{ p'{}^0_i p^0_i \ul{p}'{}^0_i \ul{p}^0_i } \Big] \slashed{M}_{p_1 p_2 \ul{p}_1\ul{p}_2 \to p'_1 p'_2 \ul{p}'_1 \ul{p}'_2} \\
& =\Big[\prod_{i}\sqrt{ (\Lambda p'_i)^0 (\Lambda p_i)^0 (\Lambda \ul{p}'_i)^0 (\Lambda \ul{p}_i)^0} \Big]
 \slashed{M}_{\Lambda p_1 \Lambda p_2 \Lambda \ul{p}_1 \Lambda \ul{p}_2 \to \Lambda p'_1 \Lambda p'_2 \Lambda \ul{p}'_1 \Lambda \ul{p}'_2}.
\end{split}
\end{align}
Hence, $\Big[\prod_{i}\sqrt{ p'{}^0_i p^0_i \ul{p}'{}^0_i \ul{p}^0_i } \Big] \slashed{M}_{p_1 p_2 \ul{p}_1\ul{p}_2 \to p'_1 p'_2 \ul{p}'_1 \ul{p}'_2}$ is Lorentz invariant under 
$p_i \rightarrow \Lambda p_i$, $p'_i \rightarrow \Lambda p'_i$, $\ul{p}_i \rightarrow \Lambda \ul{p}_i$ and $\ul{p}'_i \rightarrow \Lambda \ul{p}'_i$. This is again a straightforward generalization of the usual $S$-matrix result, and it also seems to hold for the scattering supermatrix computed for our CQ Yukawa model in Sec.~\ref{subsec:rscattering}. On the other hand, with a non-trivial translation $a^\mu \neq 0$ and $\Lambda^\mu{}_\nu = \delta^\mu_\nu$, Eq.~\eqref{app:slashedM_inv} gives 
\be
\slashed{M}_{p_1 p_2 \ul{p}_1\ul{p}_2 \to p'_1 p'_2 \ul{p}'_1 \ul{p}'_2} = \exp \left[ i\sum_{i=1,2}  ( p'_i - p_i-\ul{p}'_i + \ul{p}_i )^\mu  a_\mu \right] \slashed{M}_{p_1 p_2 \ul{p}_1\ul{p}_2 \to p'_1 p'_2 \ul{p}'_1 \ul{p}'_2}. 
\ee
Notice that this involves both the bra and ket variables. In the usual $S$-matrix constraint, the exponential is a product of two exponentials, one enforcing $\delta^4(p_1' + p_2' - p_1 - p_2)$ and the other enforcing $\delta^4(\ul{p}_1' + \ul{p}_2' - \ul{p}_1 - \ul{p}_2)$. In the more general case shown here, there are also more general delta function combinations allowed, such as those in the CQ terms of Eqs.~\eqref{eq:P-NR} or \eqref{eq:P-R}, which mix the bra and ket variables.


\section{Conservation of probability}
\label{app-probability}

In this appendix we show how to check that the total probability is conserved in this model, order by order in perturbation theory. The condition we need is that the final state density matrix has unit trace. In terms of the scattering supermatrix elements, the final state is
\be
\rho_f = \sum_{\beta \ul{\beta}} \sum_{\alpha \ul{\alpha}} \slashed{M}_{\alpha \ul{\alpha} \to \beta \ul{\beta}} \rho_{0,\alpha \ul{\alpha}} \ket{\beta} \bra{\ul{\beta}},
\ee
where $\rho_0$ is the initial state. Normalization of the final state requires
\be
1 = \tr \rho_f = \sum_{\beta} \sum_{\alpha \ul{\alpha}} \slashed{M}_{\alpha \ul{\alpha} \to \beta \beta} \rho_{0,\alpha \ul{\alpha}}.
\ee
We assume this initial state has unit trace, 
\be
1 = \tr \rho_0 = \sum_{\alpha} \rho_{0,\alpha \alpha}
\ee
in which case our requirement on the scattering elements becomes
\be
\label{eq:cons-prob}
\sum_{\beta} \slashed{M}_{\alpha \ul{\alpha} \to \beta \beta} = \delta_{\alpha \ul{\alpha}}.
\ee
Eq. \eqref{eq:cons-prob} is our elementary requirement on the scattering process which ensures that total probability is conserved. 

Perturbatively, we can be a little more explicit. In perturbation theory we can write $\slashed{M}_{\alpha \ul{\alpha} \to \beta \ul{\beta}} = \delta_{\alpha \beta} \delta_{\ul{\alpha} \ul{\beta}} + \slashed{T}_{\alpha \ul{\alpha} \to \beta \ul{\beta}}$, where the first term represents no scattering and the $\slashed{T}$ is the open systems version of the usual scattering $T$ matrix (e.g., the generalization of $S = 1 + i T$ in $S$-matrix theory). In this setting, Eq. \eqref{eq:cons-prob} becomes simply
\be
\label{eq:cons-prob-2}
\sum_{\beta} \slashed{T}_{\alpha \ul{\alpha} \to \beta \beta} = 0.
\ee
This is the open systems analogue of the optical theorem. 

In our Yukawa model, we can check Eq. \eqref{eq:cons-prob-2} order by order in $\lambda$. The first non-trivial check in $2 \to 2$ scattering is at $\Oi(\lambda^2)$. In terms of scattering probabilities, only forward scattering contributes at this order, but note that Eq. \eqref{eq:cons-prob-2} is more general. It is a statement about amplitudes for superpositions $\ket{\alpha} \bra{\ul{\alpha}}$, i.e., involves the coherence of the state. More explicitly, Eq. \eqref{eq:cons-prob-2} reads
\be
\int d^3\mb{p}_1' d^3\mb{p}_2' \slashed{T}_{\mb{p}_1 \mb{p}_2 \ul{\mb{p}}_1 \ul{\mb{p}}_2 \to \mb{p}_1' \mb{p}_2' \mb{p}_1' \mb{p}_2'} = 0.
\ee
Following the same logic as in Sec. \ref{sbsc:NR}, we have processes involving $M_{F_1}$ and $M_{F_2}$. Concretely we have, for the general process,
\begin{align}
\begin{split}
& \slashed{T}_{\mb{p}_1 \mb{p}_2 \ul{\mb{p}}_1 \ul{\mb{p}}_2 \to \mb{p}_1' \mb{p}_2' \ul{\mb{p}}_1' \ul{\mb{p}}_2'} \\
& = M_{F_1}(\mb{p}_1 \mb{p}_2 \to \mb{p}_1' \mb{p}_2') \delta^3(\ul{\mb{p}}_1 - \ul{\mb{p}}_1') \delta^3(\ul{\mb{p}}_2 - \ul{\mb{p}}_2') + \delta^3(\mb{p}_1 - \mb{p}_1') \delta^3(\mb{p}_2 - \mb{p}_2') M_{F_1^*}(\ul{\mb{p}}_1 \ul{\mb{p}}_2 \to \ul{\mb{p}}_1' \ul{\mb{p}}_2')  \\
& + \delta^3(\ul{\mb{p}}_2 - \ul{\mb{p}}_2') M_{F_2}(\mb{p}_1 \ul{\mb{p}}_2 \to \mb{p}_1' \ul{\mb{p}}_2')  \delta^3(\ul{\mb{p}}_1 - \ul{\mb{p}}_1') + \delta^3(\ul{\mb{p}}_1 - \ul{\mb{p}}_1') M_{F_2}(\mb{p}_2 \ul{\mb{p}}_1 \to \mb{p}_2' \ul{\mb{p}}_1')  \delta^3(\ul{\mb{p}}_2 - \ul{\mb{p}}_2'). 
\end{split}
\end{align}
These are the terms on the right hand side of Eq.~\eqref{eq:slashedM}, except for the trivial no-scattering term. Now we set $\mb{p}_1' = \ul{\mb{p}}_1'$, $\mb{p}_2' = \ul{\mb{p}}_2'$, and integrate over the final states, to compare to Eq.~\eqref{eq:cons-prob-2}. The result is
\begin{align}
\begin{split}
&\int d^3\mb{p}_1' d^3\mb{p}_2' \slashed{T}_{\mb{p}_1 \mb{p}_2 \ul{\mb{p}}_1 \ul{\mb{p}}_2 \to \mb{p}_1' \mb{p}_2' \mb{p}_1' \mb{p}_2'} 
\\
&= M_{F_1}(\mb{p}_1 \mb{p}_2 \to \ul{\mb{p}}_1 \ul{\mb{p}}_2)  + M_{F^*_1}(\ul{\mb{p}}_1 \ul{\mb{p}}_2 \to \mb{p}_1 \mb{p}_2)
+ M_{F_2}(\mb{p}_1 \ul{\mb{p}}_2 \to \ul{\mb{p}}_1 \mb{p}_2) + M_{F_2}(\mb{p}_2 \ul{\mb{p}}_1 \to \ul{\mb{p}}_2 \mb{p}_1)
\\
&=-\delta^4(p_1 + p_2 - \ul{p}_1 - \ul{p}_2) \frac{\lambda^2}{(2\pi)^2 m_1 m_2} \left[ \frac{i}{\mb{k}^2 + m_{\phi}^2} + \frac{D_2}{(\mb{k}^2 + m_{\phi}^2)^2} +  D_0 \right]_{\mb{k} = \ul{\mb{p}}_1 - \mb{p}_1}
\\
&
-\delta^4(\ul{p}_1 + \ul{p}_2 - p_1 - p_2) \frac{\lambda^2}{(2\pi)^2 m_1 m_2} \left[ \frac{-i}{\mb{k}^2 + m_{\phi}^2} + \frac{D_2}{(\mb{k}^2 + m_{\phi}^2)^2} +  D_0 \right]_{\mb{k} = \mb{p}_1 - \ul{\mb{p}}_1}
\\
&
+ \delta^4( p_1 - \ul{p}_2 -\ul{p}_1 + p_2 ) \frac{\lambda^2}{(2\pi)^2 m_1 m_2} \left[ \frac{D_2}{(\mb{k}^2 + m_{\phi}^2)^2} +  D_0 \right]_{\mb{k} = \ul{\mb{p}}_1 -\mb{p}_1}
\\
&
+ \delta^4( p_2 - \ul{p}_1 -\ul{p}_2 + p_1 ) \frac{\lambda^2}{(2\pi)^2 m_1 m_2} \left[ \frac{D_2}{(\mb{k}^2 + m_{\phi}^2)^2} +  D_0 \right]_{\mb{k} = \ul{\mb{p}}_2 -\mb{p}_2}
\\
&=0,
\end{split}
\end{align}
where $p^\mu_1 + p^\mu_2 = \ul{p}^\mu_1+\ul{p}^\mu_2$ holds for each term and note that the delta function of $M_{F_2}$ is different from that of $M_{F_1}$. This is because $M_{F_2}$ bridges the ket line and the bra line and because the flow of momentum for each line is opposite to each other. Hence at least to $\Oi(\lambda^2)$, the trace is preserved under CQ evolution. 

It would be very interesting to generalize this to $\Oi(\lambda^4)$, but this would require the analysis of loop integrals and we leave it to future work. We also note that even at $\Oi(\lambda^2)$, we have simplified our analysis by ignoring self-energy terms. We have also checked that including self-energy diagrams at $\Oi(\lambda^2)$ does not change the conclusion that probability is conserved. The proof is somewhat tedious so we omit it here.

\end{document}